\documentclass[manuscript]{aastex}

\newcommand{\pare}[1]{\left( #1 \right)}

\newcommand{\delr}{\frac{\partial}{\partial r}}
\newcommand{\delt}{\frac{\partial}{\partial t}}
\newcommand{\rinv}{\frac{1}{r}}
\newcommand{\OmegaK}{\Omega_{\rm K}}

\newcommand{\cs}{c_{\rm s}}
\newcommand{\massp}{m_{\rm p}}
\newcommand{\mn}{m_{\rm n}}
\newcommand{\me}{m_{\rm e}}
\newcommand{\epth}{\epsilon_{\rm TP}}
\newcommand{\Epth}{E_{\rm TP}}

\newcommand{\Uhe}{U_{\rm HEP}}
\newcommand{\pth}{p_{\rm TP}}
\newcommand{\Pth}{P_{\rm TP}}
\newcommand{\Ptot}{P_{\rm tot}}
\newcommand{\ptot}{p_{\rm tot}}
\newcommand{\phe}{p_{\rm HEP}}
\newcommand{\Phe}{P_{\rm HEP}}
\newcommand{\fvis}{f_{\rm vis}}
\newcommand{\fv}{f_{\rm v}}
\newcommand{\fcomp}{f_{\rm comp}}
\newcommand{\fc}{f_{\rm c}}
\newcommand{\Qvis}{Q_{\rm vis}}
\newcommand{\Qcompth}{Q_{\rm V,TP}}
\newcommand{\Qcomphe}{Q_{\rm V,HEP}}
\newcommand{\Qdiff}{Q_{\rm diff}}
\newcommand{\gammainj}{\gamma_{\rm inj}}
\newcommand{\gammam}{\gamma_{\rm m}}
\newcommand{\gamman}{\gamma_{\rm n}}
\newcommand{\gammath}{\gamma_{\rm TP}}
\newcommand{\gammahe}{\gamma_{\rm HEP}}

\newcommand{\nhe}{n_{\rm HEP}}
\newcommand{\nth}{n_{\rm TP}}

\newcommand{\npn}{\dot n_{\rm p\rightarrow n}}
\newcommand{\Nhe}{N_{\rm HEP}}
\newcommand{\Ninj}{\dot N_{\rm inj}}
\newcommand{\Nesc}{\dot N_{\rm esc}}
\newcommand{\Ndiff}{\dot N_{\rm diff}}
\newcommand{\Nsink}{\dot N_{\rm sink}}

\newcommand{\Qinj}{Q_{\rm inj}}
\newcommand{\Qsink}{Q_{\rm sink}}
\newcommand{\Qesc}{Q_{\rm esc}}
\newcommand{\tdiff}{t_{\rm diff}}
\newcommand{\tfall}{t_{\rm fall}}
\newcommand{\tesc}{t_{\rm esc}}
\newcommand{\tn}{t_{\rm n}}
\newcommand{\sigmat}{\sigma_{\rm T}}
\newcommand{\sigmapp}{\sigma_{\rm pp}}
\newcommand{\Cdiff}{C_{\rm diff}}
\newcommand{\rg}{r_{\rm g}}
\newcommand{\rs}{r_{\rm s}}
\newcommand{\rout}{r_{\rm out}}
\newcommand{\rin}{r_{\rm in}}
\newcommand{\rcrit}{r_{\rm crit}}
\newcommand{\mbh}{M_{\rm BH}}

\newcommand{\MdEdd}{\dot M_{\rm Edd}}

\newcommand{\LEdd}{L_{\rm Edd}}
\newcommand{\Ljet}{L_{\rm jet}}
\newcommand{\mjet}{\dot M_{\rm jet}}
\newcommand{\Ln}{L_{\rm n}}
\newcommand{\Lp}{L_{\rm p}}

\newcommand{\Ekin}{\epsilon_{\rm kin}}
\newcommand{\ppn}{P_{\rm p \rightarrow n}}
\newcommand{\tpe}{t_{\rm p-e}}

\newcommand{\taues}{\tau_{\rm es}}
\newcommand{\Tp}{T_{\rm p}}
\newcommand{\Te}{T_{\rm e}}

\slugcomment{Not to appear in Nonlearned J., 45.}

\shorttitle{Effects of HEPs on the Accretion Flows}
\shortauthors{Kimura, Toma, and Takahara}

\begin{document}


\title{EFFECTS OF HIGH-ENERGY PARTICLES ON ACCRETION FLOWS ONTO A SUPERMASSIVE BLACK HOLE}

%% USE \author, \affil, and the \and command to format

\author{Shigeo S. Kimura \altaffilmark{1}, Kenji Toma\altaffilmark{2,3}, and Fumio Takahara\altaffilmark{1}}
\altaffiltext{1}{Department of Earth and Space Science, Graduate School of Science, Osaka University, Toyonaka 560-0043, Japan}
\altaffiltext{2}{Astronomical Institute, Tohoku University, Sendai 980-8578, Japan}
\altaffiltext{3}{Frontier Research Institute for Interdisciplinary Sciences, Tohoku University, Sendai 980-8578, Japan}

%\author{C. D. Biemesderfer\altaffilmark{4,5}}
%\affil{National Optical Astronomy Observatories, Tucson, AZ 85719}
\email{kimura@vega.ess.sci.osaka-u.ac.jp}


\begin{abstract}
We study effects of high-energy particles on the accretion flows onto a supermassive black hole 
and luminosities of escaping particles such as protons, neutrons, gamma-rays, and neutrinos. 
We formulate a one-dimensional model of the two-component accretion flow 
consisting of thermal particles and high-energy particles, 
supposing that some fraction of the released energy is converted to the acceleration of the high-energy particles. 
The thermal component is governed by fluid dynamics 
while the high-energy particles obey the moment equations of the diffusion-convection equation. 
By solving the time evolution of these equations, 
we obtain advection dominated flows as the steady state solutions. 
Effects of the high-energy particles on the flow structures turn out to be small 
even if the pressure of the high-energy particles dominates over the thermal pressure. 
For a model in which the escaping protons take away almost all the released energy, 
the high-energy particles have large influence enough to make the flow have the Keplerian angular velocity at the inner region. 
We calculate the luminosities of the escaping particles for these steady solutions. 
The escaping particles can extract the energy from about $10^{-4}\dot M c^2$ to $10^{-2}\dot M c^2$, 
where $\dot M$ is the mass accretion rates. 
The luminosities of the escaping particles depend on the parameters 
such as the injection Lorentz factors, the mass accretion rates, and the diffusion coefficients. 
We also discuss some implications on the relativistic jet production by the escaping particles.
\end{abstract}





\section{INTRODUCTION} \label{sec:intro}

Active galactic nuclei (AGNs) are considered to emit high-luminosity radiation 
through accretion onto a supermassive black hole (SMBH). 
Many types of solutions of the steady accretion flow around a compact object have been found \citep[e.g.][]{sha73,abr88,bla99,yua01,oda07}. 
The advection dominated accretion flow (ADAF) is a solution 
realized when the mass accretion rates are sufficiently smaller than the Eddington accretion rates \citep{nar94}. 
Many calculations about the global structure of ADAF were performed in late 1990s \citep[e.g.][]{nar95,che97,man97}. 
ADAF is so hot and tenuous that the plasma in this flow becomes collisionless, 
which allows the particles in the flow to have a non-thermal distribution \citep{mah97b}. 
Previously, such particle acceleration has been discussed 
in the models involving shocks which may exist in the accretion flow \citep{le04,le05,bec08,bec11} 
and in the stochastic acceleration in the corona region above the disk \citep[e.g.][]{kat91,der96,sub99}. 

In this paper, we consider that particle acceleration occurs in the bulk of disk matter.
In the accretion flow, the magneto-rotational instability (MRI) plays an important role in transporting angular momentum. 
Strongly turbulent magnetic fields arise due to MRI and their stress transports the angular momentum \citep{bal91,san04}. 
Recently, numerical simulations of MRI in collisionless plasma have been performed, 
and high-energy protons are shown to be generated by magnetic reconnection induced by MRI \citep{riq12,hos13}. 
These high-energy protons are expected to interact with thermal protons and generate neutrons and pions, 
\begin{eqnarray}
p + p \rightarrow p + p + \pi^0 + X, \label{eq:pppp}\\
p + p \rightarrow p + n + \pi^+ + X,  \label{eq:pppn}
\end{eqnarray}
where X represents multiple pions. 
While thermal protons are confined by the strong turbulent magnetic fields, 
the neutrons can escape from the accretion flow because of the charge neutrality 
if their life times are longer than their escape times \citep{beg90}. 
A $\pi^0$ decays into two photons, and a $\pi^+$ decays into three neutrinos and a positron. 
Since photons and neutrinos are not trapped by the magnetic fields, 
they will also escape from the accretion flows. 
In addition, high-energy protons can have much larger mean free paths than the thermal protons 
and can escape from the flow through their diffusive motions. 
Thus, it is possible to extract energy from the accretion flows through the high-energy particles.

ADAF is also considered to be related with the formation of outflows and relativistic jets \citep{nar94,bec08}. 
However, the production mechanism of the jets is not well understood. 
If the luminosity of the jet, $\Ljet$, originates from the gravitational energy of the accreting materials, 
the condition $\Ljet < \eta\dot M c^2$ should be satisfied, 
where $\dot M$ is the accretion rate onto a SMBH and $\eta$ is the energy release efficiency.
Since $\Ljet = \Gamma \mjet c^2$, where $\Gamma\sim 10-100$ is the Lorentz factors of the jet 
and $\mjet$ is the mass loading rate to the jet, we have $\mjet \ll \dot M$. 
This means that the mechanisms concentrating the gravitational energy on a small fraction of the materials are necessary. 
The escape of the high-energy particles may be one of such mechanisms \citep[e.g.][]{le04,tom12}. 
This point also motivates us to investigate the luminosity of the escaping materials from the accretion flows. 

The energy extraction through escaping particles may affect the dynamical structure of the accretion flow. 
The high-energy particles also affect the pressure in the flow. 
While some studies consider high-energy particles for predicting photon spectra from ADAFs \citep[e.g.][]{mah97a,nie13}, 
few study the dynamical feedbacks of the high-energy particles to the accretion flows, which we study in this paper. 
We formulate one-dimensional, vertically integrated equations of the accretion flow 
including high-energy particles in Section \ref{sec:formulation}. 
Numerical results are shown in Section \ref{sec:result}. 
We discuss implications of the results and future directions of the investigation in Section \ref{sec:discussion}, 
and Section \ref{sec:summary} is devoted to the summary.

\section{FORMULATION} \label{sec:formulation}

We consider a steady accretion flow that consists of thermal and non-thermal particles.  
The thermal particles (TPs) obey the fluid equations, 
while the high-energy particles (HEPs) are described by the diffusion-convection equation \citep[e.g.][]{dru83,jon90}. 
We assume that the radiation from TPs is inefficient and ignore effects of the electron component. 
We use the cylindrical coordinate ($r, \phi, z$) and the vertically integrated equations for simplicity. 
In addition, we assume the axial symmetry. 
Under these assumptions, we treat the accretion flow as a one-dimensional problem. 
% We use the cylindrical coordinate ($r, \phi, z$) and assume the axial symmetry. 
% In addition, we use the vertically integrated equations for simplicity. 
% For a variable $a'$ such as density or velocity, its integrated variable $A$ are related to $a'$ as 
% $ A =  \int a' dz = 2H a$, where $a$ is the vertically averaged variable of $a'$. 
% %For a variable $b'$ per unit mass such as specific angular momentum, we use the vertically averaged values $b = \int b' dz/(2H)$. 
% We formulate the basic equations by using the integrated and averaged variables 
% except the velocity of vertical component (see Appendix  \ref{app:comp} for its treatment).
% Under these assumptions, we treat the accretion flow as a one-dimensional problem. 
% }

\subsection{Thermal Component}

For TPs, we assume that density, radial velocity, and angular momentum are constant for the vertical direction, 
and use vertically-integrated pressure and vertically-averaged internal energy for calculation. 
We include effects of vertical velocity for compressional heating. 
The mass and angular momentum conservations of TPs are represented as \citep[e.g.][]{pri81}
\begin{eqnarray}
  \frac{\partial \Sigma}{\partial t}+\frac{1}{r}\delr(r \Sigma v_r)=0, \label{eq:eoc} \\
  \delt(\Sigma l_z) +\frac{1}{r} \frac{\partial} {\partial r}(rv_r\Sigma l_z)
  =\frac{1}{r}\delr\left(r^3\Sigma\nu \frac{\partial\Omega}{\partial r} \right), \label{eq:eomphi} 
\end{eqnarray}
where $\Sigma$ is the surface density,  $v_r$ is the radial velocity, $l_z$ is the specific angular momentum, 
$\Omega=l_z/r^2$ is the angular velocity, and $\nu$ is the kinetic viscosity. 
We use the standard alpha prescription expressed as $\nu = \alpha \cs H$, 
where $\cs$ and $H$ are the effective sound speed and the scale height, respectively \citep{sha73}.  
We assume that the inertia of TPs is much larger than that of HEPs. 
This assumption allows us to write equations (\ref{eq:eoc}) and (\ref{eq:eomphi}) 
without any sink terms due to interchange between TP and HEP. 

The radial momentum conservation is represented as 
\begin{eqnarray}
   \delt(\rho v_r) +\frac{1}{r} \delr(r\rho v_r^2)
  +\frac{\partial}{\partial z}(\rho v_zv_r) 
  = - \frac{\partial \ptot}{\partial r}+
  \rho r\Omega^2- \rho\frac{\partial \Phi}{\partial r}, 
  \label{eq:eomr}  
\end{eqnarray}
where $\rho = \Sigma/(2H)$ is the density of TPs and $\ptot$ is the total pressure. 
We assume that $\rho$ is constant for the vertical direction. 
We use the pseudo-Newtonian potential represented as 
\begin{equation}
  \Phi = -\frac{GM}{\sqrt{r^2+z^2}-\rs},
\end{equation}
where $\rs\equiv 2GM/c^2$ is the Schwarzschild radius \citep{pac80}. 
Expanding $\Phi$ with the condition $z/r \ll 1$ and neglecting terms with $o((z/r)^3)$, 
we integrate equation (\ref{eq:eomr}) as 
\begin{eqnarray}
   \delt (\Sigma v_r) + \rinv\delr (r\Sigma v_r^2) 
  = -\frac{\partial \Ptot}{\partial r} 
  +\Sigma r(\Omega^2-\OmegaK^2)
  - \OmegaK\frac{d \OmegaK}{dr} \frac{\Sigma H^2}{3}, 
  \label{eq:eomr2}
\end{eqnarray}
where $\Ptot=\int \ptot dz$ is the integrated total pressure and 
\begin{equation}
  \OmegaK=\sqrt{\frac{GM}{r}}\frac{1}{r-\rs} 
\end{equation}
is the Keplerian angular velocity. 
The last term of equation (\ref{eq:eomr2}) accounts for the $z$ dependence of the radial component 
of the gravitational force \citep[cf.][]{mat84}. 
The integrated total pressure is represented as 
\begin{equation}
   \Ptot=\Pth+\Phe+P_B,
\end{equation}
where $\Pth$, $\Phe$, and $P_B$ are the integrated pressures of TPs, 
HEPs, and the magnetic fields, respectively. 

We reduce the vertical equation of motion to that of the hydrostatic equilibrium by neglecting the advection term, 
\begin{equation}
 H\approx\frac \cs \OmegaK  ,\label{eq:height} 
\end{equation}
where $\cs=\sqrt{\Ptot/\Sigma}$ is the effective sound speed. 

The energy conservation is
\begin{eqnarray}
  \frac{\partial }{\partial t}(\rho\epth) 
  + \frac{1}{r} \delr (rv_r\rho\epth) 
  + \frac{\partial}{\partial z}(v_z\rho\epth)  
  =-\pth \mbox{\boldmath $\nabla\cdot v$} +q_+,
  \label{eq:epth0}
\end{eqnarray}
where $\epth$ is the specific internal energy of TPs, 
$\pth$ is the pressure of the TPs, 
and $q_+$ is the viscous heating rate per unit volume for TPs. 
After some algebra, equation (\ref{eq:epth0}) is integrated and written as
\begin{equation}
   \delt( \Sigma \Epth) + \frac{1}{r} \delr \left(rv_r\Sigma\Epth\right) = Q_+, \label{eq:epth}
\end{equation}
where $\Epth = \int \epth dz/(2H)$ is the vertically averaged specific internal energy of TPs, 
%integrated pressure of the TP $\Pth$, 
and $Q_+$ is the total heating rate per unit area for TPs.
%In previous studies, the entropy conservation is often used to describe energy balance \citep[e.g.][]{nar97}. 
%In the entropy conservation equation, the compressional heating term does not appear explicitly. 
%We use the energy conservation in order to compare $\Qcompth$ to $Q_+$ explicitly. 
We will describe $Q_+$, which includes the viscous dissipation and the compressional heating, in subsection \ref{sec:energy}. 

The equation of state for TPs is written as 
\begin{equation}
   \pth  = (\gammath-1)\rho\epth, \label{eq:eosth0}
\end{equation}
where $\gammath$ is the specific heat ratio of TPs. 
Integrating this equation with the assumption that $\rho$ is independent of $z$,
we obtain the relation between $\Pth$ and $\Epth$ as 
\begin{equation}
   \Pth  = (\gammath-1)\Sigma\Epth. \label{eq:eosth}
\end{equation}
We set $\gammath=5/3$ because TPs are assumed to be non-relativistic. 
We assume turbulent magnetic fields induced by TPs in the accretion flows. 
The integrated magnetic pressure $P_B$ is estimated 
with the assumption that the plasma beta is constant, i.e
\begin{equation}
 P_B=\Pth/\beta. 
\end{equation}
%%Note that in this formulation, the magnetic fields do not behave as the relativistic gas. 
%%due to the assumption that plasma beta is constant. 
Some previous studies consider that magnetic fields behave as a relativistic gas 
and include the magnetic component in their energy equation \citep[e.g.][]{esi97}. 
However, since we hardly understand a proper description of magnetic fields, 
we simply assume that magnetic pressure is proportional to the thermal pressure 
and do not include the magnetic component in equation (\ref{eq:epth}). 
The vertically-averaged strength of the magnetic fields $B$ is defined as 
\begin{equation}
 B=\sqrt{8\pi p_B}=\sqrt{4\pi P_B/H}, 
\end{equation}
where we use $P_B = \int p_B dz = 2 H p_B $ and $p_B = B^2/(8\pi)$. 
Under this assumption, the magnetic fields do not behave as a relativistic gas. 
We assume B is constant for the vertical direction.

\subsection{High-Energy Component}

In this paper, we assume that HEPs are relativistic and regard their energy and momentum as identical. 
HEPs obey the diffusion convection equation \citep[e.g.][]{dru83,jon90} 
\begin{equation}
  \delt f + \mbox{\boldmath $v \cdot \nabla$} f = \mbox{\boldmath $\nabla$} \cdot \left( \kappa_p \mbox{\boldmath $\nabla$} f\right) 
  + \frac{\mbox{\boldmath $\nabla\cdot v$}}{3} p \frac{\partial f}{\partial p} 
  + \dot f_{\rm inj} - \dot f_{\rm sink},
  \label{eq:diffconv}
\end{equation}
where $f(t, {\bf r}, p)$ is the distribution function of HEPs, 
$\kappa_p$ is the diffusion coefficient, 
and $p$ is the momentum of HEPs. 
We add the terms $\dot f_{\rm inj}$ and $\dot f_{\rm sink}$ that describe the injection and sink, respectively. 
The sink term, added in this equation symbolically, includes effects of the neutron escape and pion production. 
We suppose that the magnetic reconnection and/or the second-order Fermi process act as the injection term. 

As a first step study, instead of solving the distribution function $f$, 
we only solve the number and energy densities, $\Nhe$ and $\Uhe$, in this paper. 
%In order to simplify the problem, we solve moment equations of the equation (\ref{eq:diffconv}) \citep[see][]{le05}. 
We define the number density and the energy density of HEPs per unit area as 
\begin{eqnarray}
 \Nhe=4\pi\int_{-\infty}^\infty dz\int_0^\infty dp p^2f,\\
 \Uhe=4\pi\int_{-\infty}^\infty dz \int_0^\infty dp p^2  f pc, 
\end{eqnarray}
respectively. 
%In this study, we do not treat the spectrum of HEPs, 
%and thus, we use mean Lorentz factor when necessary. 
We can treat the mean Lorentz factor as
\begin{equation}
  \gammam\equiv \frac{\Uhe}{\massp c^2 \Nhe },
\end{equation}
where $\massp$ is the proton mass. 
Taking the appropriate moments of equation (\ref{eq:diffconv}) and integrating over the vertical direction, 
we obtain the equations of number and energy densities of HEPs as 
\begin{eqnarray}
  \frac{\partial \Nhe}{\partial t} +  \rinv \delr ( r v_r \Nhe)  = 
  \rinv \delr \left( r \kappa \frac{\partial \Nhe}{\partial r} \right) -
  \Ndiff + \Ninj - \Nsink,  \label{eq:Nhe}  \\
  \frac{\partial \Uhe}{\partial t}  +  \rinv \delr ( r v_r \Uhe)  =  \Qcomphe + 
  \rinv \delr \left( r\kappa \frac{\partial \Uhe}{\partial r} \right) -
  \Qdiff + \Qinj - \Qsink,      \label{eq:Uhe} 
\end{eqnarray}
respectively. 
We have used the averaged diffusion coefficient $\kappa$, 
the injection terms $\Ninj$ and $\Qinj$, 
the sink terms $\Nsink$ and $\Qsink$, 
and the escaping rates of HEPs through vertical diffusion 
$\Ndiff=\Nhe/\tdiff$ and $\Qdiff=\Uhe/\tdiff$, 
where $\tdiff=H^2/\kappa$ is the vertical diffusion time. 
Equation (\ref{eq:Uhe}) has the compressional heating term 
\begin{equation}
 \Qcomphe = -\frac{\Phe}{rH}\delr (rHv_r) , 
\end{equation}
where we define the integrated pressure of HEPs as 
\begin{equation}
 \Phe=4\pi\int_{-\infty}^\infty dz \int_0^\infty dp p^2 f \frac{c p}{3} . 
\end{equation}
See Appendix \ref{app:comp} for the treatment of the compressional heating term. 
Since we assume that HEPs are relativistic, the relation between $\Uhe$ and $\Phe$ is given as 
\begin{equation}
  \Uhe = 3\Phe.
\end{equation}
This relation implies that $\gammahe=4/3$, where $\gammahe$ is the specific heat ratio of HEPs. 
We describe $\kappa$ as 
\begin{equation}
   \kappa = \frac{1}{3}c \lambda 
  = \frac{\Cdiff c \rg }{3} .
\end{equation}
We represent the mean free path as $\lambda=\Cdiff \rg$, 
where $\Cdiff$ is a parameter that represents difference from the Bohm diffusion, 
and $\rg=\gammam \massp c^2/(eB)$ is the gyro radius. 
In the actual situation, the diffusion coefficient depends on the Lorentz factor of the particles 
because the particles with higher energies have larger mean free paths. 
As a first step study, however, we do not treat the spectrum of HEPs 
but simplify the situation by taking the moments of the distribution function $f$. 
In the same spirit, we use $\gammam$ when we estimate $\rg$.

\subsection{Energy Dissipation and Energy Loss} \label{sec:energy}

HEPs affect the dynamical structure of the flow through the pressure term and energy extraction. 
In this subsection, we summarize the internal energy injected into or extracted from the accretion flows. 
In this paper, we assume that the injection rates into HEPs are related to the heating rates of TPs. 
TPs are heated by the viscous dissipation rates, 
\begin{equation}
   \Qvis= \Sigma \nu \left(r\frac{\partial \Omega}{\partial r}\right)^2, 
\end{equation}
and the compressional heating rates,
\begin{equation}
 \Qcompth = -\frac{\Pth}{rH}\delr (rHv_r). \label{eq:Qcompth}
\end{equation} 
See Appendix \ref{app:comp} for derivation of equation (\ref{eq:Qcompth}). 
Since the turbulent viscosity is expected to induce the dissipation in the accretion flows, 
it is considered that some fraction of the dissipated power is expended to inject HEPs by the second-order Fermi acceleration. 
The compression of the turbulent magnetic fields is likely to induce the magnetic reconnection, 
so that HEPs are expected to be generated by consuming some fraction of the compressional heating energy. 
Thus, we assume that the fraction $\fvis$ of $\Qvis$ is injected into HEPs, 
and the remaining fraction $(1-\fvis)$ goes into TPs. 
Similarly, the fraction $\fcomp$ of $\Qcompth$ goes into HEPs, 
and the other $(1-\fcomp)$ heats up TPs, i.e.
\begin{equation}
  \Qinj = \fvis \Qvis + \fcomp \Qcompth
\end{equation}
and
\begin{equation}
 Q_+=(1-\fvis)\Qvis + (1-\fcomp) \Qcompth. 
\end{equation}
As described above, we ignore the spectrum of HEPs and only use the mean Lorentz factor. 
In the same manner, we assume mono-energetic injection everywhere. 
Using the Lorentz factor at injection, $\gammainj$, 
the injection term for the number density of HEPs is represented as
\begin{equation}
  \Ninj = \frac{\Qinj}{(\gammainj-1)\massp c^2}. 
\end{equation}

%In order to avoid the inflow of HEPs at the outer boundary, 
%We will explain why we introduce the critical radius in the next subsection. 
%We treat $\gammainj$, $\rcrit$, and $\fv$ as parameters. 

The interactions between HEPs and TPs extract the energy and particles from the flow. 
Since we ignore radiation processes of TPs, 
we do not consider the background photon fields. 
This treatment allow us to neglect photomeson production, $p\gamma \rightarrow p \pi^0\ or\ n \pi^+ $, 
and consider only proton-proton collisions ($pp$ collisions). 
When the reactions (\ref{eq:pppp}) and (\ref{eq:pppn}) occur, 
pions are produced, and high-energy protons or neutrons lose their energies. 
The pions decay into photons, neutrinos, electrons, and positrons as \citep{beg90} 
\begin{eqnarray}
\pi^0 \rightarrow 2\gamma, \label{eq:pi0}\\ 
\pi^+ \rightarrow e^+ + 3\nu, \label{eq:pi+}\\
 \pi^- \rightarrow e^- + 3\nu. \label{eq:pi-}
\end{eqnarray}
Since we consider tenuous accretion flows with the optical depth for electron scattering $\taues\lesssim1$, 
the neutrinos and photons can escape directly from the flows, 
and the high-energy electrons and positrons are considered to emit radiation and lose their energy rapidly. 
Thus, the flows lose their energy by pion production through $pp$ collisions. 
Using the inelasticity of this reaction, $K_\pi$, 
we estimate the energy loss rates by pion production as 
\begin{equation}
  Q_\pi  = \int K_\pi \Ekin \nhe \nth \sigmapp cdz
   = \frac{K_\pi \Ekin \Nhe \Sigma \sigmapp c}{2\massp H}, 
\end{equation}
where $\nhe=\Nhe/(2H)$ is the number density of HEPs, $\nth=\Sigma/(2\massp H)$ is the number density of the TPs, 
$\Ekin=(\gammam-1)\massp c^2$ is the mean kinetic energy of HEPs, and 
\begin{equation}
\sigmapp = 30\left[0.95 + 0.06 \ln \pare{\frac{\Ekin}{1\rm GeV}}\right]\rm mb
\end{equation}
is the cross section for $pp$ collisions \citep{aha00}. 
We assume that the number density of HEPs is uniform for the vertical direction 
when estimating the $pp$ collision rate.

Neutrons are also produced by $pp$ collisions. 
The formation rates of relativistic neutrons in unit volume are estimated as 
\begin{equation}
   \npn  = \frac 1 2 \ppn \nhe \nth \sigma_{\rm pp} c
\end{equation}
where $\ppn$ is the probability for neutron formation per interaction. 
The factor $1/2$ indicates that half of the neutrons are thermal. 
Neutrons may escape from the flows because of the charge neutrality, 
whereas neutrons decay into the protons when their life time has passed after their formation. 
In order that a neutron escapes from the flow, 
its escape time $\tesc$ has to be shorter than its life time $\tn=887\gamman $sec, 
where $\gamman$ is the Lorentz factor of the escaping neutron. 
Ignoring escape of thermal neutrons from the flows since most of thermal neutrons satisfy $\tn \ll \tesc$, 
we write the neutron escape rates as 
\begin{equation}
  \Nesc = \int \npn \exp\left(-\frac{\tesc}{\tn }\right)dz
\end{equation}
With the approximation that all neutrons move along the vertical direction, 
we write the escaping time as $\tesc=(H-z)/c$, 
and the neutron escape rates are evaluated as 
\begin{equation}
  \Nesc = \npn c \tn \left\{1-\exp\left( -\frac{2H}{c \tn}\right) \right\}. 
\end{equation}
On neutron production, some fraction of energy is carried away by pions, 
and the Lorentz factor of escaping neutrons satisfies the condition of
\begin{equation}
  (\gamman-1)\mn c^2 = (1-K_\pi)(\gammam - 1)\massp c^2, 
\end{equation}
where $\mn$ is the mass of a neutron. 
We can neglect the interactions of neutrons with TPs because we consider tenuous flows \citep{beg90,tom12}. 
Using $\gamman$, we represent the energy loss rates by neutron escape as
\begin{equation}
   \Qesc = (\gamman -1)\mn c^2 \Nesc. 
\end{equation}

The sink term of equation (\ref{eq:Nhe}) is equivalent to neutron escape,
\begin{equation}
\Nsink=\Nesc .
\end{equation}
On the other hand, $\Qsink$ in equation (\ref{eq:Uhe}) includes the cooling by pion production in addition to neutron escape,
 \begin{equation}
 \Qsink=\Qesc +Q_\pi. 
\end{equation}
We set $K_\pi = 0.5$ and $\ppn = 0.5$ following to \citet{beg90}.

\subsection{Calculation Method and Conditions}

We solve the six differential equations, (\ref{eq:eoc}), (\ref{eq:eomphi}), 
(\ref{eq:eomr2}), (\ref{eq:epth}), (\ref{eq:Nhe}), and (\ref{eq:Uhe}) for $\Sigma,\ l_z,\ v_r,\ \Epth,\ \Nhe,$ and $\Uhe$. 
We calculate the time evolution of these equations until a steady state solution is realized 
rather than solve the equations with steady assumption because the former method has some advantages over the latter. 
One of the advantages is that we need not treat the singular point arising in the steady state flow equations. %due to the transonic flow. 
Another advantage is that unstable solutions are not realized. % in the case for calculating the time evolution. 
In order to solve the fluid equations, 
we use a method of finite differences with a time-explicit solution procedure similar in methodology to the ZEUS code 
with the von Neumann \& Richtmyer artificial viscosity \citep{neu50,sto92}. 
The equations of HEPs are solved by using the fully-implicit method \citep{pre92}. 
We determine the time step so that the CFL condition is safely satisfied (the safety factor $C_0=0.1$). 
The number of the grid points is $N=256$, and the grids are uniformly divided in the logarithmic space. 
We calculate some models with $N=128$ and find that the results are unchanged by the number of grids. 

The initial conditions are unimportant because the system forgets them by the time when a steady state solution is realized. 
We set the initial conditions as follows, 
\begin{eqnarray}
\Sigma = -\frac{\dot M }{2\pi r v_r},\\
l_z = 0.9\OmegaK r^2,\\
v_r = v_{r,0}r^{-1}, \\
\Epth = -0.5\Phi,\\
\Nhe = 0.0 ,\\
\Uhe = 0.0,
\end{eqnarray}
where $\dot M$ is the mass accretion rates, and $v_{r,0}$ is determined to be smoothly connected at the outer boundary. 
The boundary conditions do not strongly affect the solutions when we choose sufficiently large $\rout$. 
We assume that there is a rotationally supported flow at the outer boundary $r=\rout$, 
i.e., we set the outer boundary of TPs as 
\begin{eqnarray}
 \Sigma = -\frac{\dot M }{2\pi r v_r},\\
l_z = 0.9\OmegaK \rout^2, \\
v_r = -\frac{3\nu}{2\rout}, \\
\Epth  = -0.5\Phi. 
\end{eqnarray}
These boundary conditions make the viscous dissipation rates large, 
which are expected to induce the large injection rates. 
We confirm that the results are almost unchanged if we set a slowly rotating outer boundary, such as $l_z=0.3 \OmegaK \rout^2$. 
For HEPs, we set the outflow boundary condition. 
Under this condition, the inflow of HEPs at the outer boundary is prohibited
so that HEPs that diffuse out from the outer boundary do not return into the calculated region. 
In this study, we assume that HEPs are accelerated only within the critical radius, 
i.e., the allocation factors are given as 
\begin{eqnarray}
 \fvis=\left\{ \begin{array}{ll}
\fv & ( r < \rcrit) \\
0 & (r > \rcrit) \\
\end{array} \right. ,\\
 \fcomp=\left\{ \begin{array}{ll}
\fc & ( r < \rcrit) \\
0 & (r > \rcrit) \\
\end{array} \right. .
\end{eqnarray}
%%where $\rcrit$ is the critical radius within which particle acceleration occurs, 
%%and $\fv$ is the allocation parameter.
We treat $\rcrit$, $\fv, $and $\fc$ as parameters. 
We set the free boundary conditions for the inner boundary at $r=\rin$ 
because the flow should be supersonic at the vicinity of the black hole. 
All the variables satisfy the condition $\partial /\partial r=0$ at $r=\rin$.

%%%%%%%%%%%%%%%%%%%%%%%%%%%%%%%%%%%%%%%%%%%%%%%%
%%Here, we should describe why we employ this condition.%%
%%%%%%%%%%%%%%%%%%%%%%%%%%%%%%%%%%%%%%%%%%%%%%%%

\section{CALCULATION RESULTS}\label{sec:result}

In our formulation, there are several free parameters, 
such as the diffusion parameter $\Cdiff$, the allocation parameters $\fv$ and $\fc$, 
the critical radius $\rcrit$, and the injection Lorentz factor $\gammainj$. 
Since it is too complex to study with all the parameters varying, 
we fix the parameters $\mbh,\ \rout,\ \rin,$ and $\rcrit$, which are tabulated in Table \ref{tab:fix}. 
We choose $\rout=150\rs$ in order to shorten the calculation time. 
Effects of HEPs are expected to be large as $\rcrit$ is large, 
and we use $\rcrit = 100\rs$.
We calculate with the other parameters tabulated in Table \ref{tab:models}.
The group A consists of the models without HEPs ($\fv=\fc=0$). 
We compare the results of the groups A with the previous global solutions of ADAF 
in order to confirm validity and consistency of our formulation and method. 
By comparing results among the groups B, C, and D, 
we investigate effects of the ways how to inject HEPs. 
Injection rates in the group B, C, and D are respectively proportional to 
the viscous dissipation rates, the compressional heating rates, and the total heating rates. 
We consider a model E1 in which HEPs take away almost all energy. 
The dynamical structure of this model is very different from the structures without HEPs.

\subsection{Dynamical structure of flows without High-Energy Particles}

\begin{figure}
   \centering
   \epsscale{1.0}
   \plotone{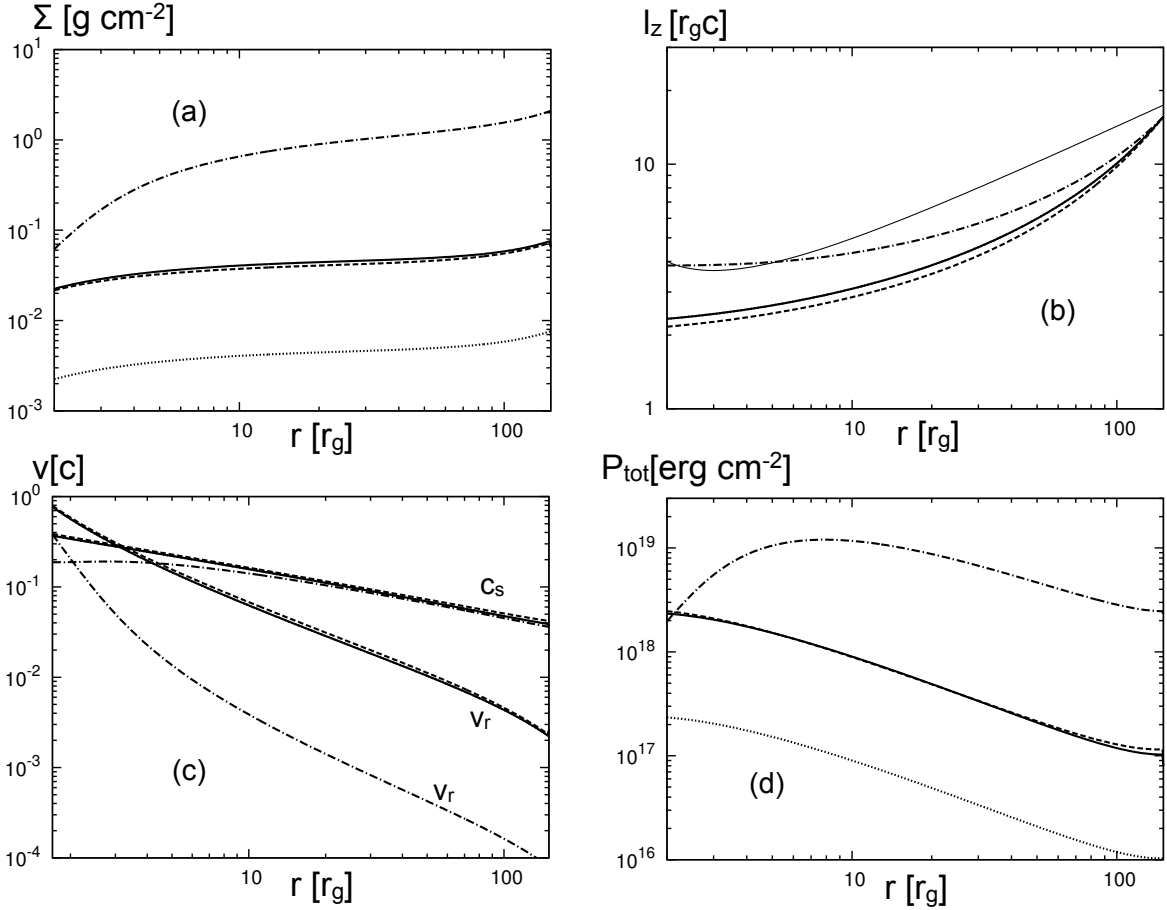}
   \caption{Radial distributions of (a) the surface density $\Sigma$, (b) the specific angular momentum $l_z$, 
   (c) the radial velocity $v_r$ and the effective sound speed $\cs$, and (d) the integrated total pressure $\Ptot$
   for the group A. The solid, dashed, dotted, and dot-dashed lines are for the models A1 (reference), A2 (small $\beta$), 
   A3 (small $\dot M$), and A4 (small $\alpha$), respectively. 
   The thin solid line in (b) represents the Keplerian angular momentum. 
   \label{fig:f0}}
   \end{figure}

We show the results of the group A for which there are no HEP. 
These results correspond to ADAF models with no radiative cooling. 
A1 is a reference model, A2 is a model with strong magnetic fields, 
A3 with a small mass accretion rate, and A4 with a small $\alpha$ parameter. 
Figure \ref{fig:f0} shows the radial distributions of (a) the surface density $\Sigma$, (b) the specific angular momentum $l_z$, 
(c) the radial velocity $v_r$ and effective sound speed $\cs$, and (d) the integrated total pressure $\Ptot$ for the group A. 
%A1 (solid, reference), A2 (dashed, small $\beta$), A3 (dotted, small $\dot M$), and A4 (dot-dashed, small $\alpha$). 
From (c) of Figure \ref{fig:f0}, we find that transonic solutions are realized in all models 
by solving time evolution of a system of fluid equations. 
The sonic radii of our solutions are between $2\rs$ and $4\rs$, 
which are consistent with previous global solutions of ADAF \citep{che97,nak97,nar97}. 

Comparing A1 ($\beta=10$, solid lines) with A2 ($\beta=3$, dashed lines), 
we found that the strength of the magnetic pressure scarcely affects the dynamical structure. 
The dashed lines in Figure \ref{fig:f0} almost overlap with the solid lines. 
This feature is consistent with the previous solutions \citep{nak97}. 
%Since $P_B\propto \beta$, A2 has three times larger magnetic pressure than A1. 
%However, the magnetic pressure is not important for the dynamical structure 
%because we consider the thermal pressure dominated flow. 
%If we consider magnetic pressure dominated model, 
%we should include an equation describing the evolution of magnetic fields \citep[e.g.][]{oda07}.  
The mass accretion rate affects the surface density and the total pressure. 
The surface density is proportional to $\dot M$, 
and the total pressure $\Ptot \propto \Sigma \propto \dot M$. 
We can see this feature in panels (a) and (d) by comparing A3 ($\dot M=0.001$, dotted lines) with A1 ($\dot M=0.01$). 
However, the mass accretion rate has very little influence on the structure of $l_z$, $v_r$, and $\cs$. 
In panels (b) and (c), the dotted lines completely overlap with the solid lines. 
These dependences on the mass accretion rate are common features of ADAF solutions \citep{nar94,kat08}. 

The $\alpha$ parameter strongly affects the dynamical structure of the flows. 
For A4 ($\alpha=0.003$), $v_r$ and $\Sigma$ are respectively small and large,
while $\cs$ is not very different, compared with the reference model A1 ($\alpha=0.1$).
This makes the sonic radius smaller. 
The small $\alpha$ parameter makes the transport of the angular momentum inefficient, 
and the flow rotates super Keplerian in $r\simeq 3-4 \rs$. 
To realize a transonic solution, the radial velocity rapidly increases as $r\rightarrow \rs$. 
This makes the surface density rapidly decrease while $\cs$ is almost constant at the inner region $r\lesssim7\rs$. 
This causes the integrated pressure to decrease rapidly there. 
Thus, the integrated pressure has the maximum at $r\simeq 7 \rs$. 
These features are also seen in the previous solutions \citep{che97,nak97,nar97}. 
Therefore, our solutions are consistent with numerical solutions found in other studies.

\begin{figure}
   \centering
   \epsscale{0.7}
   \plotone{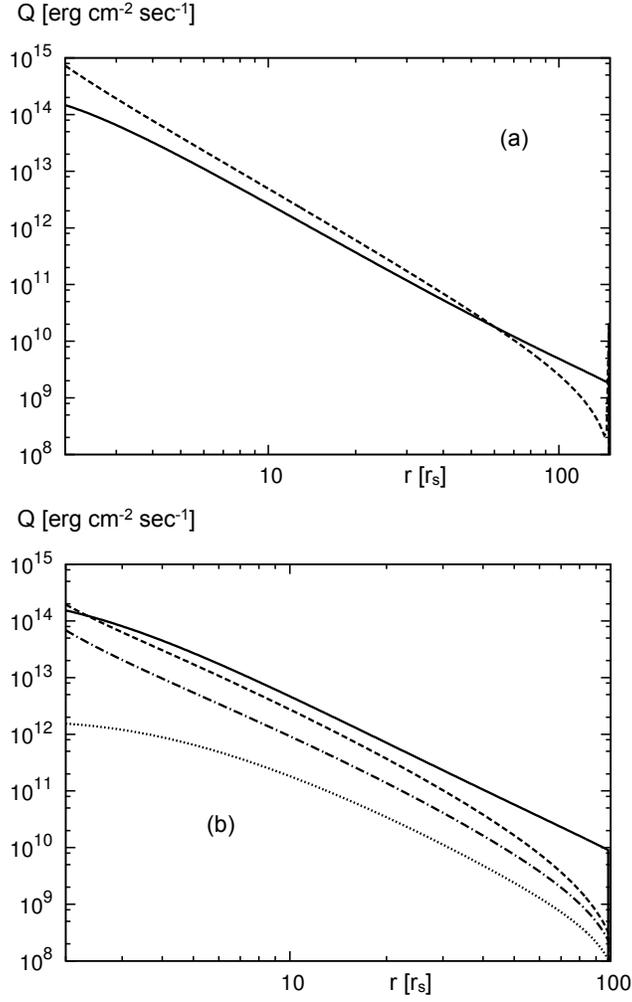}
   \caption{(a) Radial distributions of heating rates for A1. 
   The solid and dashed lines show the viscous heating rate and the compressional heating rate, respectively. 
   (b) Radial distributions of the heating and cooling rates for HEPs in D3. 
   The solid, dashed, dotted, and dot-dashed lines show 
   the injection rate, the compressional heating rate, the cooling rate by $pp$ collisions, and the proton escaping rate, respectively.  
   \label{fig:heat}}
   \end{figure}

We check the energy balance of the flow. 
Figure \ref{fig:heat} (a) represents the heating rates for A1. 
The solid and dashed lines show the viscous dissipation and compressional heating rates, respectively. 
It is seen that the compressional heating is dominant in the inner region ($r\lesssim60\rs$) 
while the viscous dissipation is larger than the compressional heating in the outer region ($r\gtrsim 60\rs$). 
However, the compressional heating rate is at most eight times larger than the viscous dissipation rate at the innermost region. 
Both the compressional heating and the viscous dissipation are important to heat up TPs in this model. 
In previous papers, the energy balance was discussed by using the entropy. 
In that viewpoint, the compressional heating is included in the advection term of the entropy \citep[cf.][]{nar94,nar97}, 
and what determines the internal energy has not been explicitly discussed.

\subsection{Dynamical structure of flows with High-Energy Particles}

\begin{figure}
   \centering
   \epsscale{1.0}
   \plotone{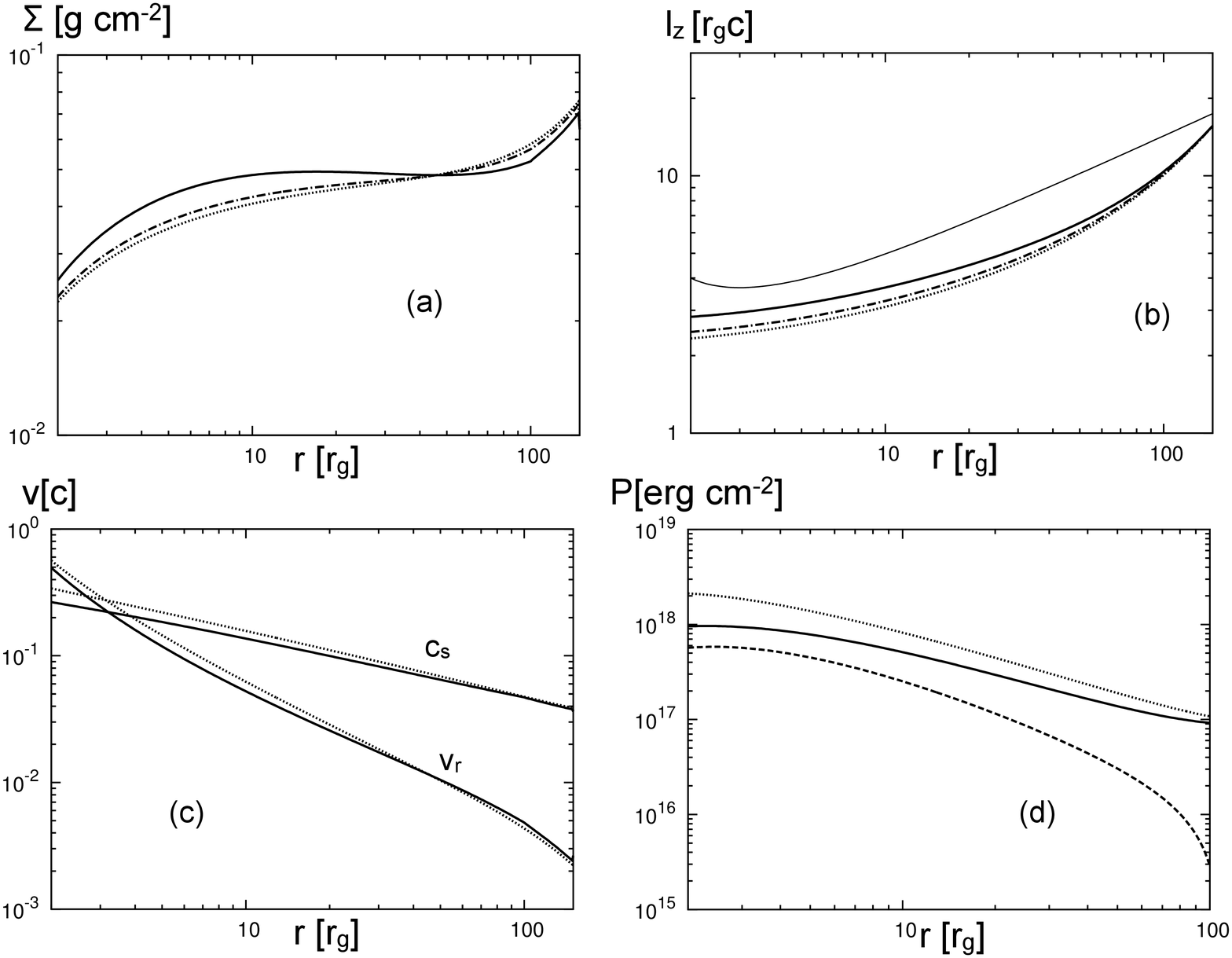}
   \caption{Radial distributions of (a) the surface density $\Sigma$, (b) the specific angular momentum $l_z$, 
   (c) the radial velocity $v_r$ and the effective sound speed $\cs$, for the group B. 
   The solid, dot-dashed, and dotted lines are for B2 ($\fv=0.9$), B1 ($\fv=0.3$), and A1 (no HEPs for reference), respectively. 
   The thin solid line in (b) represents the Keplerian angular momentum. 
   (d) Radial distributions of the integrated pressure for B2. The solid and dashed lines represent $\Pth$ and $\Phe$, respectively. 
   The dotted line depicts $\Pth$ for A1 (no HEPs) for reference. 
   \label{fig:fc0fa}}
  \end{figure}

In this subsection, we show the results of the models including HEPs.
First, we compare the results with different injection models. 
The group B consists of the models with $\fv \ne 0$ and $\fc=0$, in which injection rates are related only to the viscous dissipation. 
Figure \ref{fig:fc0fa} shows the radial distributions of (a) the surface density $\Sigma$, (b) the specific angular momentum $l_z$, 
and (c) the radial velocity $v_r$ and the effective sound speed $\cs$. 
From panels (a) and (b), the surface density and angular momentum distributions of B1 ($\fv=0.3$) are almost same as A1 ($\fv=\fc=0$). 
The surface density and the specific angular momentum of B2 ($\fv=0.9$) are a few tens of percent larger than those of A1. 
Similarly, $v_r$ and $\cs$ of B2 are a few tens of percent smaller than those of A1. 
Panel (d) shows the radial distributions of the integrated pressure for B2, 
from which we found that $\Pth$ is about twice larger than $\Phe$. 
Even with $\fv=0.9$, $\Pth$ always dominates over $\Phe$. 
This is due to the efficient compressional heating, 
which dominates over the viscous dissipation in the inner region. 
It is found that HEPs have little influence on the dynamical structure for the models in the group B 
because the total injected energy is not so large compared with the total energy that heats up TPs. 
The thermal pressure for B2 is nearly half of that for A1. 
This is because some fraction of the dissipation energy is expended for injection of HEPs. 
Note that the compressional heating rate of B2 is about twice smaller than that of A1 
since the compressional heating rates are proportional to $\Pth$. 

\begin{figure}
   \centering
   \epsscale{1.0}
   \plotone{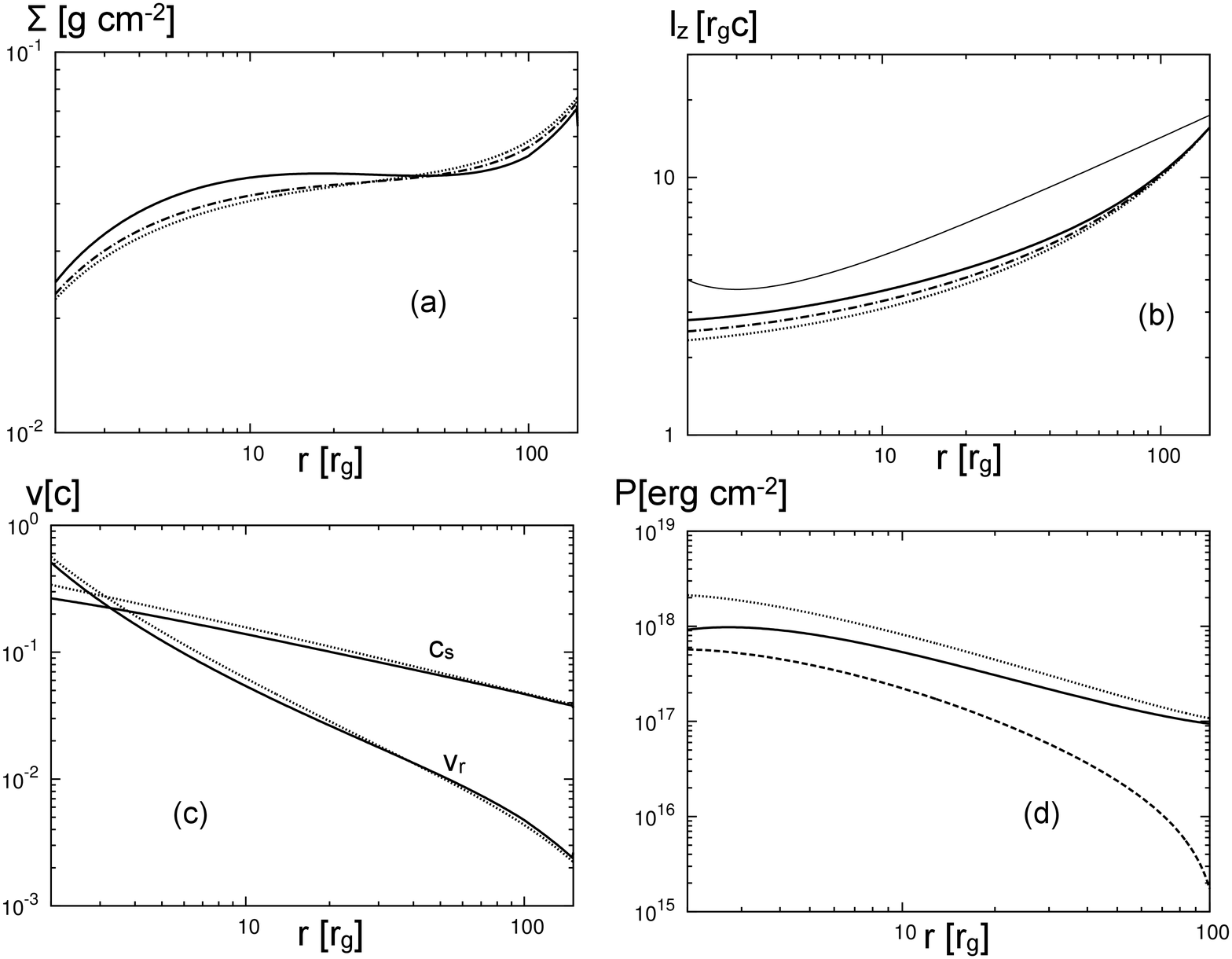}
   \caption{Radial distributions of (a) the surface density $\Sigma$, (b) the specific angular momentum $l_z$, 
   (c) the radial velocity $v_r$ and the effective sound speed $\cs$, for the group C. 
   The solid, dot-dashed, and dotted lines are for C2 ($\fc=0.9$), C1 ($\fc=0.3$), and A1 (no HEPs for reference), respectively. 
   The thin solid line in (b) represents the Keplerian angular momentum. 
   (d) Radial distributions of the integrated pressure for C2. The solid and dashed lines represent $\Pth$ and $\Phe$, respectively. 
   The dotted line depicts $\Pth$ for A1 (no HEPs) for reference. 
   \label{fig:fa0fc}}
  \end{figure}

Since the compressional heating dominates over the viscous dissipation at the inner region, 
it is worth to investigate the effects of injection related only to the compressional heating. 
The group C consists of the models with $\fv=0$ and $\fc\ne 0$, 
in which the injection rates are related only to the compressional heating. 
Figure \ref{fig:fa0fc} shows the radial structures of the solutions of the group C.
They are quite similar to those of the group B. 
From (a) and (b) of Figure \ref{fig:fa0fc}, 
it is seen that the radial structure of C1 ($\fc=0.3$) is nearly the same as that of A1 ($\fv=\fc=0$). 
We found that HEPs scarcely affect the dynamical structure 
and that $\Pth > \Phe$ everywhere even for the model C2 ($\fc=0.9$). 
At $r=\rcrit$, the viscous dissipation rate is larger than the compressional heating rate, 
so that $\Phe$ in the outer region is slightly smaller than that for the group B. 
The compressional heating is expended to inject HEPs rather than to heat up TPs. 
This causes the specific internal energy of TPs to be small, 
and the angular momentum of C2 is slightly larger than that of A1 owing to inefficient transport of the angular momentum. 
This makes the viscous dissipation rate slightly larger, 
and the injection rate is smaller than the dissipation rate except for the innermost region $r\lesssim 3\rs$. 
Thus, the injection only from the compressional heating cannot energize HEPs enough to satisfy $\Phe > \Pth$. 
Although the compression does not heat up TPs in the group C, 
the slightly large dissipation rate causes the total heating rate for TPs in C2 to be nearly the same as that in B2. 
This is why the results of the group C is quite similar to those of the group B. 

\begin{figure}
   \centering
   \epsscale{1.0}
   \plotone{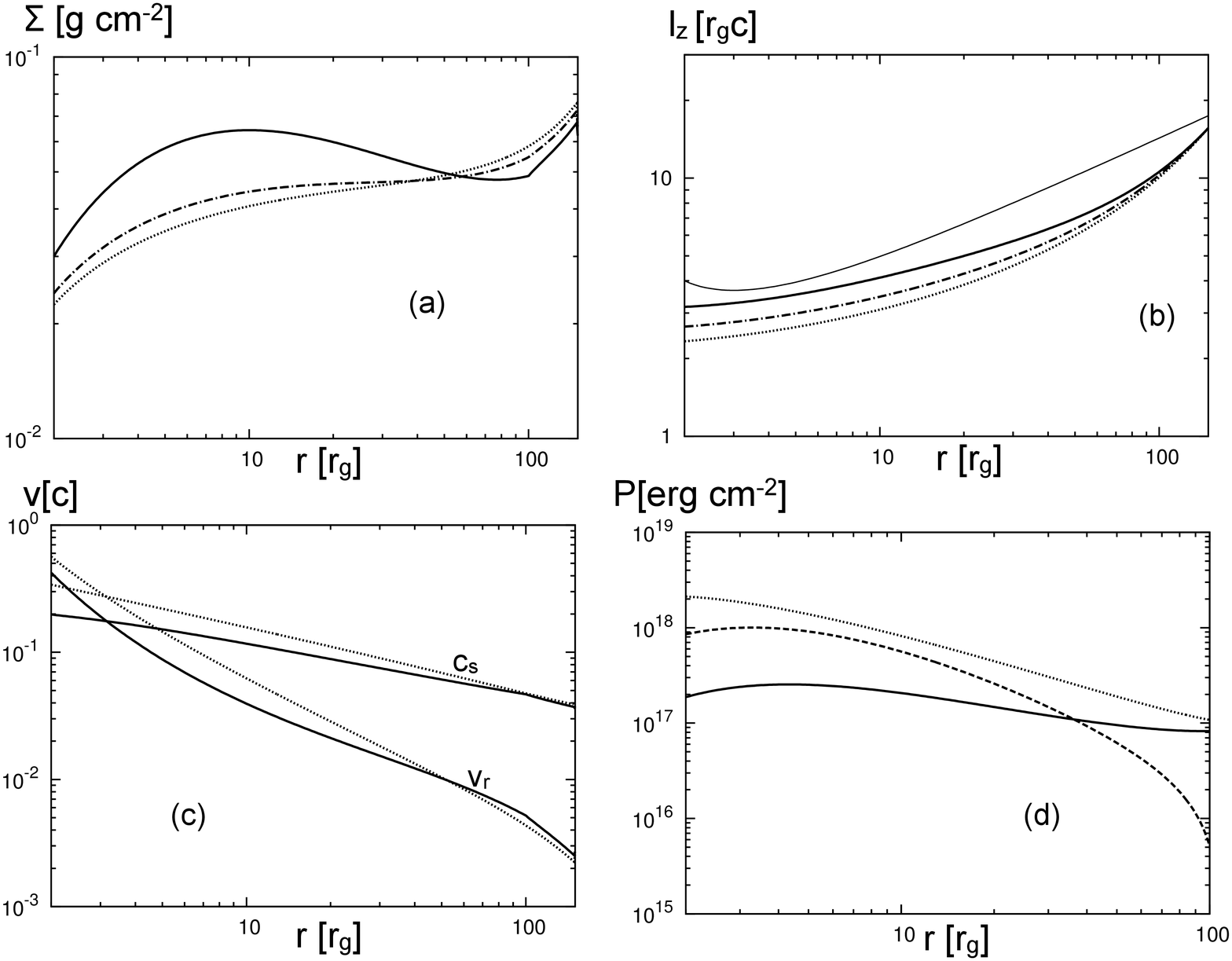}
   \caption{Radial distributions of (a) the surface density $\Sigma$, (b) the specific angular momentum $l_z$, 
   (c) the radial velocity $v_r$ and the effective sound speed $\cs$, for the group D. 
   The solid, dot-dashed, and dotted lines are for D3 ($\fv=\fc=0.9$), D1 ($\fv=\fc=0.3$), and A1 (no HEPs for reference), respectively. 
   The thin solid line in (b) represents the Keplerian angular momentum. 
   (d) Radial distributions of the integrated pressure for D3. The solid and dashed lines represent $\Pth$ and $\Phe$, respectively. 
   The dotted line depicts $\Pth$ for A1 (no HEPs) for reference. 
   \label{fig:fafc}}
  \end{figure}

The pressure of HEPs does not dominate over the thermal pressure in the groups B and C. 
This motivates us to investigate the models in which the injection rates are proportional to the total heating rates. 
The group D consists of such models with $\fv=\fc\ne 0$. 
Figure \ref{fig:fafc} shows the radial structures of the solutions of the models D1 and D3. 
From panels (a) and (b) of Figure \ref{fig:fafc}, 
we can see that the profiles of $\Sigma$ and $l_z$ of D1 ($\fv=\fc=0.3$) are nearly the same as those of A1. 
For this model, the allocation factors $\fc$ and $\fv$ are so small that $\Phe < \Pth$ is satisfied everywhere. 
From (d) of Figure \ref{fig:fafc}, we see that $\Pth < \Phe$ in $r\lesssim 40\rs$ for D3 ($\fv=\fc=0.9$).
The thermal pressure of D3 has about ten times smaller than that of A1 in $r \lesssim 10 \rs$
since almost all released energy is spent to inject HEPs.
Although $\Pth < \Phe$ is realized when HEPs are injected from both the viscous dissipation and the compressional heating, 
other variables for D3 are at most a few times larger or smaller than A1. 
Even if $\Phe > \Pth$, HEPs does not strongly affect the radial profiles of $v_r,\ \cs,\ l_z,$ and $\Sigma$. 

We explain how HEPs affect the dynamical structure of the accretion flows. 
The solutions with larger $\fv$ and/or $\fc$ have slightly larger $\Sigma$, 
larger $l_z$, smaller $v_r$, smaller $\cs$, and smaller $\Ptot$. 
As HEPs gain large fraction of released energy, the specific heat ratio of accreting materials is smaller. 
This makes $\Ptot$ small, and the angular momentum transport is inefficient. 
This causes the angular momentum to be large, and the large centrifugal force makes $v_r$ small. 
The small $v_r$ causes $\Sigma$ to be large so that the mass accretion rate is constant. 
However, the effects of the specific heat ratio cannot change the dynamical structure by an order of magnitude 
even if $\Pth < \Phe$ is satisfied. 
The flow structures are the advection dominated flows for the models in groups B, C, and D. 

\begin{figure}
   \centering
   \epsscale{1.0}
   \plotone{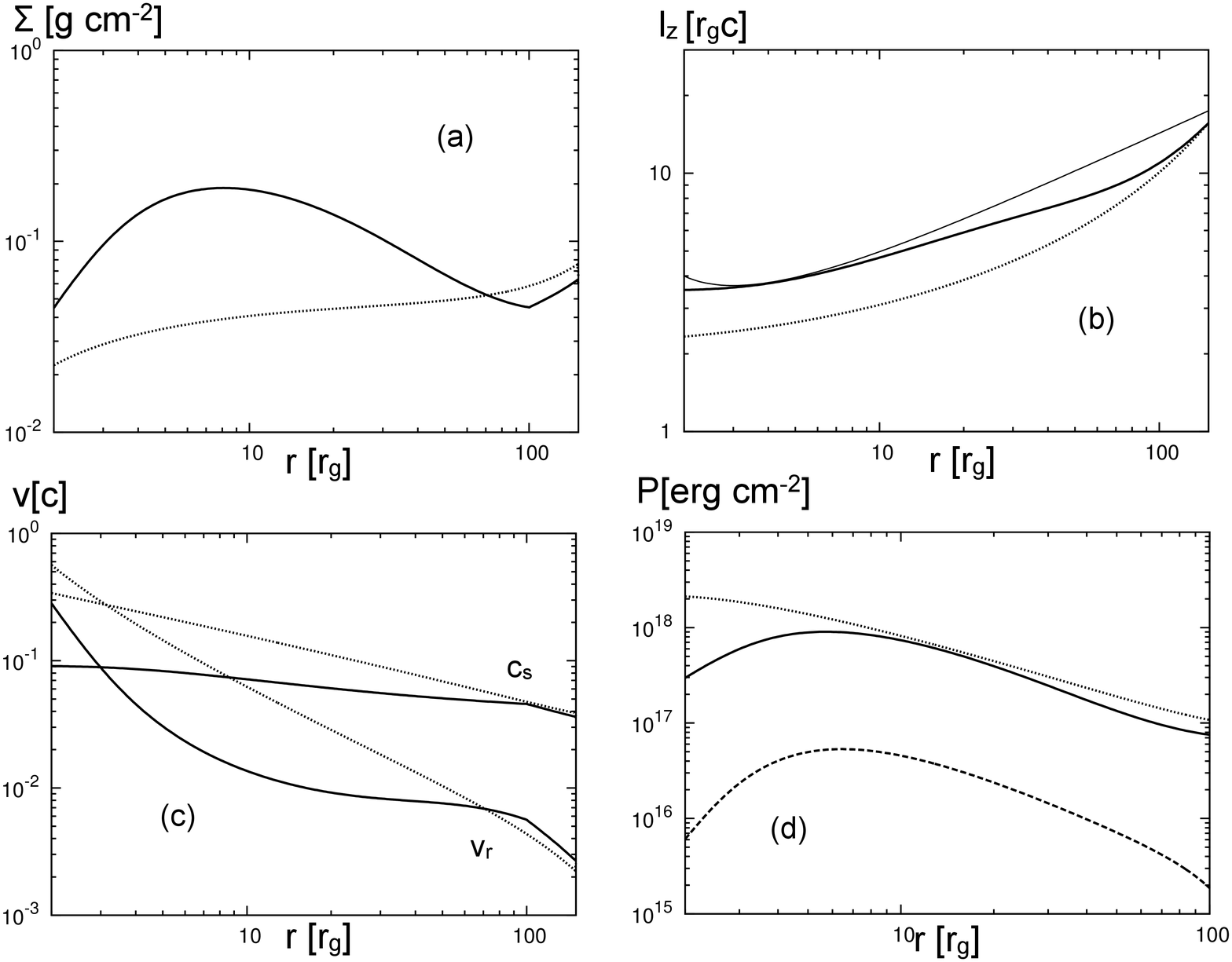}
   \caption{Radial distributions of (a) the surface density $\Sigma$, (b) the specific angular momentum $l_z$, 
   (c) the radial velocity $v_r$ and the effective sound speed $\cs$, for the model E1. 
   The solid and dotted lines are for E1 ($\fv=\fc=0.9$) and A1 (no HEPs for reference), respectively. 
   The thin solid line in (b) represents the Keplerian angular momentum. 
   (d) Radial distributions of the integrated pressure for E1. The solid and dashed lines represent $\Pth$ and $\Phe$, respectively. 
   The dotted line depicts $\Pth$ for A1 (no HEPs) for reference. 
   \label{fig:extreme}}
  \end{figure}

This ADAF structures can be changed by HEPs when they extract almost all released energy. 
We calculate the model E1 ($\fv=\fc=0.9,\ \Cdiff=10^4$) in which the accretion flow loses most of the energy by proton escape. 
Figure \ref{fig:extreme} shows the results of E1.  
In this model, the integrated pressure of HEPs is much smaller than $\Pth$ 
because the escaping protons take away almost all the injected energy. 
This makes $\cs$ small, which causes $l_z$ to be large. 
From panel (b), we can see that in $r\lesssim 8 \rs$, the flow has the Keplerian angular momentum. 
Since the centrifugal force is large owing to the large $l_z$, 
$v_r$ is small and thereby $\Sigma$ is large.
Although HEPs extract almost all energy, 
$\Pth$ for E1 is not so different from that for A1 except in $r\lesssim 5\rs$. 
This is because the increment of $\Sigma$ balances the decrement of $\cs$. 
To realize a transonic solution, 
the radial velocity is rapidly increasing in the inner region ($r\lesssim5\rs$). 
This causes the surface density to be rapidly decreasing, 
so that the integrated pressure has maximum at $r\sim 5 \rs$. 
This result indicates that an ADAF solution changes to a Keplerian thin disk 
when almost all energy is taken away from the accretion flow by HEPs, 
which is consistent with the self-similar solution obtained by \citet{nar94}. 
However, this drastic change makes this model inconsistent with the assumption of ignoring radiative cooling (see \ref{sec:ignored}). 

Next, we discuss what determines the number and energy densities of HEPs. 
Figure \ref{fig:heat} (b) indicates the heating rates and cooling rates of HEPs for D3. 
We can see that the injection rate $\Qinj$ and the compressional heating rate $\Qcomphe$ are larger than 
the cooling rate by $pp$ collisions $\Qsink$ and the diffusive escaping rate $\Qdiff$ everywhere. 
We find that $\Qcomphe \lesssim \Qinj$ 
and that $\Qcomphe$ is not so large as to make $\gammam$ much larger than $\gammainj$. 
Thus, the mean Lorentz factor is nearly the same value as the injected value, 
\begin{equation}
 \gammam \sim \gammainj. \label{eq:gamma}
\end{equation}
This condition is satisfied within a factor of two. 
This result implies that the balance between $\Qinj$ and the advection term, 
which is the second term of the left side of equation (\ref{eq:Uhe}), determines $\Uhe$ and $\Nhe$. 
Note that the dominant process energizing the HEPs is different among D1, D2, and D3. 
For D3 ($\fv=\fc=0.9$), the injection from the viscous dissipation mainly energizes HEPs 
because $\Pth$ is so small that the injection from the compressional heating is inefficient. 
On the other hand, the injection from the compressional heating is dominant for D1 ($\fv=\fc=0.3$) 
because $\Pth$ is large enough to satisfy $\Qvis < \Qcompth$. 
Both the viscous dissipation and the compressional heating make nearly the same contribution to the injection for D2 ($\fv=\fc=0.6$).

\subsection{Luminosities of Escaping Particles} \label{sec:luminosity}

We also calculate luminosities of escaping gamma-rays, neutrinos, neutrons, and protons. 
We define the luminosities as 
\begin{equation}
 L_i = \int_{\rin}^{\rout} 2\pi r Q_i dr, 
\end{equation}
where $i$ refers to the kind of escaping particles and $Q_i$ is the energy flux. 
We use $Q_{\rm n} = \Qesc$ for the neutron luminosity 
and $Q_{\rm p}=\Qdiff$ for the proton luminosity. 
%%The gamma-rays, neutrinos, and neutrons are produced by $pp$ collisions. 
For estimating the luminosity of gamma-rays and neutrinos, 
we assume that all kinds of pions produced by $pp$ collisions have the same energy, 
$Q_{\pi^j}=Q_\pi /3$, where $j=+,\ -,$ or $0$. 
Neutral pions decay into gamma-rays as equation (\ref{eq:pi0}), 
and charged pions decay into neutrinos, electrons, and positrons as equations (\ref{eq:pi+}) and (\ref{eq:pi-}).  
%$\pi^\pm \rightarrow \mu^\pm + \nu_\mu (\overline \nu_\mu),\ \mu^\pm \rightarrow e^\pm + \nu_e (\overline \nu_e) + \overline \nu_\mu(\nu_\mu)$. 
The electrons and the positrons are considered to lose most of their energies rapidly by emitting gamma-rays, 
and thus, we assume that their energies are converted to the energy of gamma-rays. 
Roughly speaking, the pion energy is equally divided among the final products \citep{beg90}. 
Under these assumptions and assuming that all of the photons and neutrinos can escape, 
$Q_\nu$ and $Q_\gamma$ are represented as
\begin{eqnarray}
 Q_\nu = \frac 3 4 Q_{\pi^+} + \frac 3 4 Q_{\pi^-} = \frac 1 2 Q_\pi, \\
 Q_\gamma = Q_{\pi^0} + \frac 1 4 Q_{\pi^+} + \frac 1 4 Q_{\pi^-} = \frac 1 2 Q_\pi. 
\end{eqnarray}
In this treatment, $Q_\gamma=Q_\nu$ is always satisfied, which leads to $L_\gamma=L_\nu$. 
When all the neutrons escape, the ratio of $\Ln$ to $L_\gamma (=L_\nu)$ 
is determined exclusively by $\ppn$ and $K_\pi$ as 
\begin{equation}
 \Ln/ L_\gamma = [\ppn (1-K_\pi)]/K_\pi. 
\end{equation}
In this model, we use $\ppn=1/2$ and $K_\pi=1/2$, so that $\Ln/L_\gamma=1/2$. 

%We show only $L_\gamma$ in this paper. 

\begin{figure}
   \centering
   \epsscale{1.0}
   \plotone{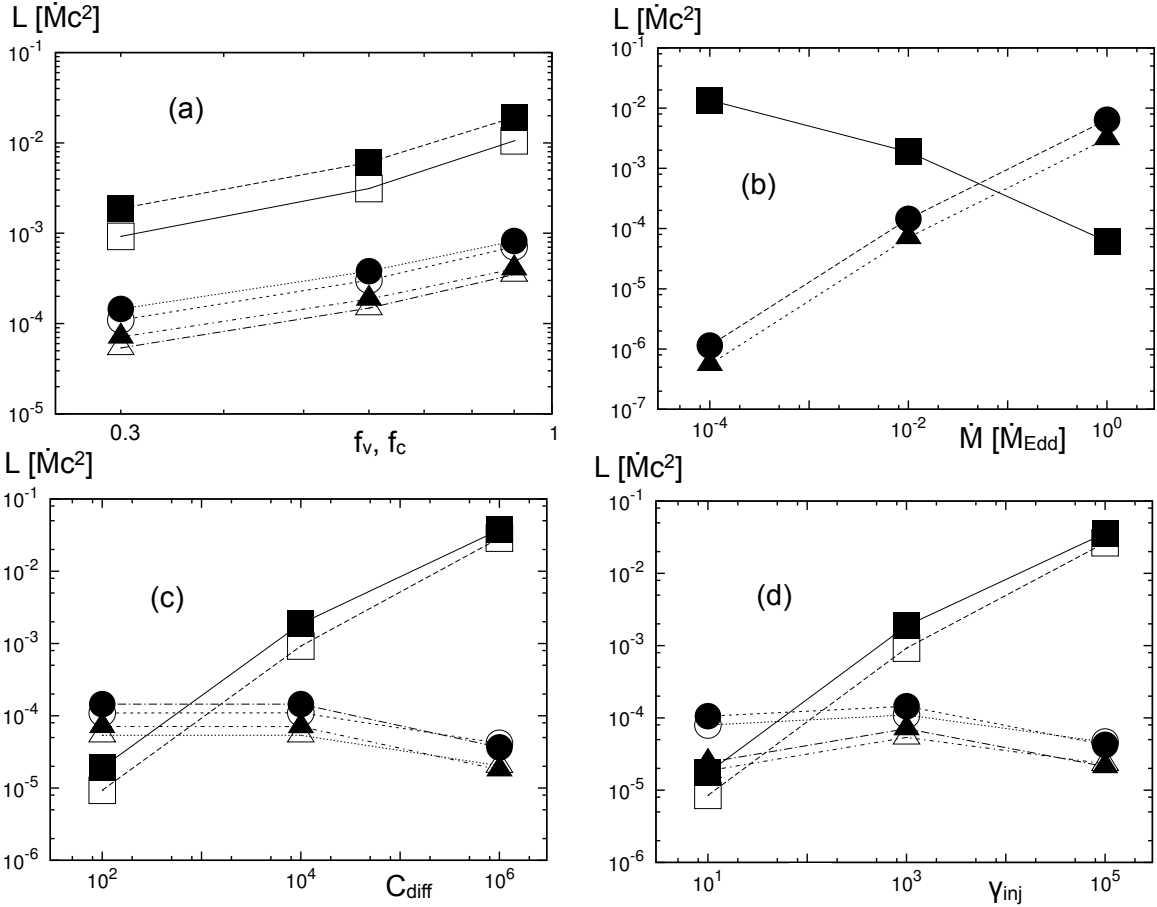}
   \caption{Luminosities of the escaping particles as the functions of parameters. 
   The squares, the circles, and the triangles denote $\Lp$, $L_\gamma=L_\nu$, and $\Ln$, respectively.
   The open and filled symbols are $\beta=3$ and $\beta=10$, respectively.    
   (a) the dependence on $\fv=\fc$. 
   This panel shows the results for D1, D2, D3, F1, F2, and F3. 
   (b) the dependence on $\dot M$. 
   This panel shows the results for D1, G1 and G2. 
   (c) the dependence on $\Cdiff$. 
   This panel shows the results for D1, H1, H2, I1, I2, and I3.
   (d) the dependence on $\gammainj$. 
   This panel shows the results for D1, J1, J2, K1, K2, and K3. 
   \label{fig:luminosity}   }
  \end{figure}

We see the parameter dependences of the luminosities of the escaping particles. 
We choose the model D1 as a reference model. 
The parameters of the models calculated additionally are tabulated in Table \ref{tab:dependence}. 
We calculate various values of $\fv=\fc$ (for the groups D and F ), $\dot M$ (for the group G), 
$\Cdiff$ (for the groups H and I), and $\gammainj$ (for the groups J and K). 
Figure \ref{fig:luminosity} shows the luminosities of protons, neutrons, and gamma-rays, $\Lp,\ \Ln,$ and $L_\gamma$. 
Panel (a) shows the luminosities as a function of the allocation parameters under the condition $\fv=\fc$, 
where we show the results of the groups D and F. 
We calculate the models in the group F in order to show effects of $\beta$. 
The luminosity of the protons is the largest of the three and reaches about $2\times10^{-2}\dot M c^2$. 
The gamma-ray or neutrino luminosity is smaller than $\Lp$ by about an order of magnitude, $L_\gamma=L_\nu \lesssim 10^{-3}\dot M c^2$. 
In this model, most of the generated neutrons can escape owing to large $\gammainj$, 
so that $\Ln/L_\gamma=1/2$. 
Large $\fv$ makes the thermal pressure small and thereby weakens the magnetic fields, 
which makes the diffusion coefficient larger. 
Thus, the dependence of $\Lp$ on $\fv$ is slightly stronger as $\fv$ is closer to unity. 
The proton luminosity is small for small $\beta$ because the strong magnetic fields prevent the protons from escaping. 
On the other hand, $\Ln$ and $L_\gamma$ are nearly independent of $\beta$ owing to their charge neutrality. 

Panel (b) represents the $\dot M$ dependence, where we show the results for D3, G1, and G2.
For the small mass accretion rate $\dot M = 10^{-4}\MdEdd$, $\Lp\sim 10^{-2}\dot M c^2$ and $\Ln\sim 10^{-6} \dot M c^2$ 
while $\Lp\sim 10^{-4}\dot M c^2$ and $\Ln\sim 3\times 10^{-3} \dot M c^2$ for the large mass accretion rate $\dot M = \MdEdd$. 
The large mass accretion rates strengthen magnetic fields and thereby decrease the diffusion coefficient 
as $\kappa\propto B^{-1}\propto \dot M^{-1/2}$. 
The large mass accretion rates also strengthen injection rates, 
which makes the energy density of HEPs larger as $\Uhe\propto\Qinj \propto \Ptot \propto \dot M$. 
Thus, roughly speaking, $\Lp \propto \kappa\Uhe \propto B\Ptot \propto \dot M^{1/2}$. 
Note that if we normalize $\Lp$ by the accretion luminosity $\dot M c^2$,  
it is a decreasing function of mass accretion rates as $\Lp/(\dot M c^2)\propto \dot M^{-1/2}$. 
The neutrons and $\gamma$-ray luminosities are nearly proportional to $\dot M^2$ 
since $L \propto \Sigma\Nhe \propto \dot M^2$. 
We can see that $\Lp > \Ln$ for $\dot M \lesssim 10^{-1}\MdEdd$ and vice versa. 

Panel (c) expresses the dependence on $\Cdiff$, where we show the results for D3 and the models in the groups H and I.
The models in the group I is different from the group H in the value of $\beta$. 
For $\Cdiff \lesssim 10^4$, the diffusive escaping rate is not so large 
that the balance of advection and injection determines the energy density of HEPs. 
In this situation,  $\Lp \propto \kappa \propto \Cdiff B \propto \Cdiff \beta^{1/2}$, 
and $\Ln$ and $L_\gamma$ are not affected by the diffusion phenomena and thereby nearly independent of $\Cdiff$ and $\beta$. 
However, for very large $\Cdiff$, the escaping rate is large enough to balance the injection rate, 
so that $\Lp$ is limited at $\Lp \sim 0.1\fv\dot M c^2$. 
Since the injection rates are nearly independent of $\beta$, 
$\Lp$ for I2 ($\beta=10,\ \Cdiff=10^6$) is nearly equal to that for J3 ($\beta=3,\ \Cdiff=10^6$). 
Efficient proton escape makes $\Nhe$ small, which decreases the collision rate. 
Thus, $\Ln$ and $L_\gamma$ with $\Cdiff=10^6$ are several times smaller than those in $\Cdiff=10^4$. 

Panel (d) depicts the $\gammainj$ dependence of luminosities, where we show the results for D3 and the models in the groups J and K. 
The models in the group K is different from the group J in the value of $\beta$. 
The $\Lp$ and $L_\gamma$ in panel (d) is quite similar to those in (c). 
For $\gammainj\lesssim 10^3$, the proton luminosity is proportional to $\gammainj$ since $\Lp \propto \kappa\propto\gammainj$. 
The gamma-ray luminosity is nearly independent of $\gammainj$. 
The number density of HEPs is inversely proportional to $\gammainj$, 
while the energy per interaction is proportional to $\gammainj$. 
Since these effects balances, $L_\gamma$ is nearly independent of $\gammainj$. 
On the other hand, the neutron luminosity with $\gammainj=10$ is a few times smaller than that with $\gammainj=10^3$. 
This is because the neutrons cannot escape from the outer region ($r\sim100\rs$) with small $\gammainj$ 
while they can escape with large $\gammainj$. 
For very large $\gammainj\sim 10^5$, 
the proton escaping rate is so large that escaping protons can extract almost all injected energy. 
This is the same situation as the case with very large $\Cdiff$. 
The proton luminosity is nearly equal to the total injection luminosity $0.1 \fv \dot M c^2$, 
and $\Ln$ and $L_\gamma$ with $\gammainj=10^5$ are several times smaller than those with $\gammainj=10^3$. 

The proton luminosity strongly depends on many uncertain parameters such as $\gammainj$ and $\Cdiff$. 
This is due to the uncertainty of diffusion and acceleration of HEPs in the accretion flows. 
On the other hand, the gamma-ray, neutrino, and neutron luminosities does not have strong dependence on such parameters. 
These luminosities strongly depend only on the mass accretion rates. 
For widely acceptable ADAF mass accretion rates ($\dot M \lesssim 10^{-2}\MdEdd$), 
these luminosities are less than about $10^{-4}\dot M c^2$. 
This value is negligibly small to change the dynamical structure from ADAF to the standard disk like structure.

We also estimate the mass escaping rates defined as 
\begin{equation}
 \dot M_i = \int_{\rin}^{\rout} 2\pi r m_i\dot N_i dr, 
\end{equation}
where $i=$p or n. 
We use $\dot N_{\rm p}=\Ndiff$ and $\dot N_{\rm n}=\Nsink$. 
Escaping protons have the Lorentz factor $\gamma_{\rm esc} \sim \gammainj$, 
and thus, we can write the mass escaping rates as 
\begin{equation}
 \dot M_{\rm p} \sim \frac{\Lp}{\gammainj c^2}. \label{eq:dmescp}
\end{equation}
The Lorentz factor of escaping neutrons is nearly half of $\gammainj$. 
The mass escaping rates of escaping neutrons are represented as 
\begin{equation}
 \dot M_{\rm n} \sim \frac{2\Ln}{\gammainj c^2}. \label{eq:dmescn}
\end{equation}
Since $\Lp \propto \gammainj$ in usual, 
$\dot M_{\rm p}$ is independent of $\gammainj$. 
On the other hand, $\dot M_{\rm n}$ is smaller as $\gammainj$ is smaller 
because $\Ln$ has the weak dependence on $\gammainj$. 
We find that $\dot M_{\rm p}\lesssim 10^{-4}\dot M$ and 
$\dot M_{\rm n}\lesssim 10^{-3}\dot M$ in our calculation. 
Since both $\dot M_{\rm n}$ and $\dot M_{\rm p}$ are sufficiently less than $\dot M$, 
the assumption that we neglect the sink term in equations (\ref{eq:eoc}) and (\ref{eq:eomphi}) is valid.

\section{DISCUSSION}\label{sec:discussion}

\subsection{Implications for Jet Production}

The observations suggest that 
Lorentz factors of the jets are typically $\Gamma\sim 10-100$, 
and that their luminosities are broadly distributed over $\Ljet \lesssim \dot M c^2$ \citep{fer11,pun11}. 
As described in Section \ref{sec:intro}, 
if the energy source of a jet is gravitational energy that is released by mass accretion, 
some mechanisms that concentrate the energy on a small fraction of mass 
are necessary in order to produce relativistic jets.  
It is likely that the gravitational energy is converted to Poynting and/or kinetic energies 
and they are injected into the polar region above the SMBH, ``the funnel'', 
where the gas is very dilute due to the centrifugal barrier. 
The most actively discussed model is the magnetically driven jet model 
investigated by the magneto-hydrodynamics simulations \citep{mck06,kom07}. 
The electromagnetic force accelerates the flow to relativistic speed, 
and it is considered that the amount of mass injected in the funnel determines the terminal Lorentz factor. 
Alternative idea is the kinetically dominated jet model, 
in which the relativistic thermal energy (i.e., random kinetic energy of particles) 
is transferred to the acceleration of the bulk flow \citep{asa07,bec11}. 
In this model, the terminal Lorentz factor is roughly equal to 
the averaged random Lorentz factor of particles.

HEPs that escape from the accretion flows are likely to inject some amount of kinetic energy and mass in the funnel, 
which is available to launch the kinetically dominated jet.  
Although the escaping particles are considered to be isotropic, 
we discuss the case with the most efficient injection 
in which all the escaping particles are injected in the funnel. 
If the accretion rate is large, the neutron luminosity is larger than the proton luminosity, 
and it amounts to $\Ln\sim 10^{-2}\dot M c^2$ for G2.
We note that what happens in the large mass accretion rate is controversial 
because the electron component is not expected to be negligible (see subsection \ref{sec:ignored}). 
For the smaller mass accretion rates, $\Ln$ is smaller since the neutron production is ineffective. 
In such situation, $\Lp$ is larger than $\Ln$ if $\gammainj$ is large, 
and the proton luminosity attains $\Lp\sim 10^{-2}\dot M c^2$ 
for the efficient escaping models H2, I3, J2, and K3 ($\Cdiff=10^6$ or $\gammainj=10^5$). 
Therefore, for AGN jets with $\Ljet\lesssim 10^{-2}\dot M c^2$, 
the energy injection by escaping particles is one of the viable mechanisms 
to launch a relativistic jet over a broad range of mass accretion rates. 
However, if there is no other mass injection except escape of HEPs, 
the terminal Lorentz factor of the jet is estimated as $\Gamma\sim \gammainj$. 
For $\gammainj=1000$, this value is too large in comparison to observed values. 

For bright AGN jets that have $\Ljet \gtrsim 10^{-2} \dot M c^2$, 
the energy injection rates by escaping HEPs are not sufficient. 
The magnetically dominated jet models are feasible for these jets. 
Although HEPs are expected to act as the source of mass injection, 
they cannot inject sufficient amount of mass in our model. 
Mass injection rates to jets are $\dot M_{\rm n}\lesssim10^{-3}\dot M$ 
for neutrons and $\dot M_{\rm p}\lesssim10^{-4}$ for protons. 
This seems too small in order to explain the bright AGN jets $\Ljet\sim \LEdd$ with $\Gamma=10-100$. 

\citet{tom12} first calculated the injection rates of mass and energy in the funnel by escaping neutrons. 
They used a power-law energy spectrum of the isotropically escaping neutrons and 
calculated the injection rates only for the neutrons that decay in the funnel, 
although they did not solve the structure of the accretion flow. 
They estimate $\Ln\lesssim 2\times 10^{-3}\dot M c^2$ and 
$\dot M_{\rm n}\lesssim 6\times 10^{-4}\dot M$. 
The total rates including the neutrons that do not decay in the funnel, i.e.,
the isotropic escaping rates, are around $\Ln\sim 0.03\dot M c^2$, which is slightly larger than those in our models. 
This is because they assume the large heating rate and the small infall timescale $\tfall\equiv r/v_r$ at the vicinity of a SMBH.
On the other hand, our model does not include the spectrum of HEPs 
that is considered to affect escaping rates of HEPs. 
In order to clarify injection rates of mass and kinetic energy, 
we should construct a more realistic model (see Subsection \ref{sec:ignored}).

\subsection{Effects of Ignored Processes}\label{sec:ignored}
In this paper, we have ignored effects of the electron component and radiation from thermal component. 
If electrons obtain large amount of thermal energy, 
they radiate the energy away by synchrotron emission and bremsstrahlung. 
Under the assumption that electrons obtain thermal energy from protons by Coulomb collisions and that electrons are non-relativistic, 
the timescale of energy transport from protons to electrons is estimated as \citep[cf.][]{spi62,tak85}
\begin{equation}
 \tpe = \sqrt{\frac \pi 2}\frac \massp \me \frac{1}{n\sigmat c \ln \Lambda}
  \pare{\frac{k\Tp}{\massp c^2}+ \frac{k\Te}{\me c^2}}^{3/2},
%\frac{4\sqrt{\pi}}{\kes c^4 \ln \Lambda} \pare{\frac \massp \me} 
%  \frac{H \cs^3}{\Sigma}, % \pare{\theta_e+\theta_p}^{3/2}, 
\end{equation}
where we use the Coulomb logarithm $\ln\Lambda$, 
the Boltzmann constant $k$, 
the proton temperature $\Tp$, 
the electron temperature $\Te$, 
and the electron mass $\me$. 
We estimate $\tpe$ under the assumption that $\Te/\me=\Tp/\massp$. 
If the energy transport time $\tpe$ is less than the infall time $\tfall$, 
the effects of the electron component should be relevant. 
At $r\sim 10\rs$, the ratio of these two timescales is roughly 
 $ \tpe /\tfall \sim10$ for $ \dot M = 0.01\MdEdd $ 
and  $ \tpe /\tfall \sim0.1$ for $ \dot M = \MdEdd $. 
Thus, for small mass accretion rates like $\dot M=0.01\MdEdd$, 
the electrons does not affect the dynamics of the flow 
%%since the condition $\tpe > \tfall$ is satisfied except around the outer boundary. 
%%the condition $\tpe < \tfall$ is satisfied except for very inner region around the sonic point, 
%%which means that 
whereas the effects of electrons should not be ignored for large mass accretion rates as $\dot M\simeq \MdEdd$.
The solutions realized in such situations are not well-understood, and thus, we do not get involved with this problem in this paper. 
If we consider large $\fv$ and $\fc$, the density is large, and $\cs$ and $v_r$ are small. 
This makes it difficult to satisfy $\tpe > \tfall$. 
For model D3 $\tpe / \tfall \sim 0.3$ at $r\sim 10\rs$, 
and for model E1, $\tpe / \tfall \sim 0.03$ at $r\sim 10\rs$
even if $\dot M = 0.01 \MdEdd$. 
Thus, when HEPs affect the dynamical structure, the electrons are also expected to play important roles on the dynamical structure. 

The ADAF solution is considered to produce not only jets but also disk winds \citep[see][]{nar94,bla99}. 
Many studies on accretion flows with the multi-dimensional simulations show that the disk winds are very common phenomena \cite[e.g.][]{mck06,ohs11}. 
The disk winds affect the mass accretion rates, angular momentum transport, and internal energy. 
Though it is important to include effects of the disk winds, 
modeling those effects in the one-dimensional model is not simple. 
The multi-dimensional study is necessary in order to understand the effects of the disk winds, and it remains as a future work.  

Turbulent magnetic fields in the accretion flow are related to the acceleration and diffusion process of HEPs. 
According to the quasi-linear theory of the wave-particle interaction, 
$\Cdiff$ is related to the strength of the turbulent magnetic fields at the scale of the resonant wave length. 
In accretion flows, turbulent magnetic fields are expected to be induced by MRI. 
Typically, injection scale of the turbulent magnetic fields, which is around the scale height of the accretion flow, 
is about 10 orders of magnitude larger than gyration scale of HEPs \citep{der96}. 
This difference between two scales is expected to make the turbulent fields very weak at the gyration scale of HEPs. 
Thus, Bohm limit that corresponds to $\Cdiff=1$ is unlikely to be suitable in the accretion flows, 
and we have used $\Cdiff=10^2,\ 10^4,\ \rm and\ 10^6$. 
We note that acceleration of HEPs is inefficient for large $\Cdiff$ 
because large $\Cdiff$ means that HEPs rarely interact with the turbulent magnetic fields. 
From the point of view of particle acceleration, 
it seems difficult to produce a large amount of HEPs by stochastic acceleration for the models with $\Cdiff=10^6$. 

% Since it is expected that $\Cdiff=10^5-10^7$ for shocks in blazars \citep{ino96}, 
% the value $\Cdiff=10^6$ is possible. 
% Large $\Cdiff$ makes $\tdiff$ short and acceleration time long as 
% \begin{equation}
%  t_{\rm acc} \sim \frac {\Cdiff \rg}{c}\frac {c^2}{v_{\rm A}^2},
% \end{equation}
% where $v_{\rm A}^2=B^2/(4\pi\rho)$ is Alfven speed. 
% At $r=10\rs$, $t_{\rm acc}/\tdiff \sim 3\times 10^{-4}$ for D3 ($\gammainj=10^3,\ \Cdiff=10^4$), 
% so that acceleration to this energy is likely to occur. 
% On the other hand, $t_{\rm acc}/\tdiff \sim 10$ for E1 ($\gammainj=10^3$, $\Cdiff=10^6$) 
% and $t_{\rm acc}/\tdiff \sim 2$ for I3 ($\gammainj=10^3$, $\Cdiff=10^4$). 
% These models are inconsistent in point of the timescales for acceleration. 

We assume monoenergetic HEPs 
in order to use the moment equations of the diffusion convection equation. 
Actually, HEPs have energy spectra that are determined 
by acceleration, escape, and cooling processes \cite[e.g.][]{der96}. 
Owing to the energy dependence of the diffusion coefficient, 
particles with larger energy are considered to escape from the flow faster than those with lower energy. 
This feature is likely to affect the luminosity and mass escaping rates of protons. 
In order to discuss the diffusive phenomena more precisely, 
we should model and solve the acceleration process with including the momentum dependence of HEPs.

\section{SUMMARY}\label{sec:summary}
We have studied the effects of high-energy particles on the accretion flow onto a supermassive black hole. 
We also calculate luminosities of escaping particles such as protons, the neutrons, the gamma-rays, and the neutrinos. 

We formulate a one-dimensional model of the two component accretion flow consisting of the thermal particles and the high-energy particles. 
The thermal component is governed by fluid dynamics, where we ignore effects of radiative cooling. 
For high-energy particles, the moment equations of the diffusion-convection equation are solved 
with accounting for coolings by pion production, neutron escape, and proton escape.
We assume that injection rates of high-energy particles are related to heating rates of thermal particles. 
We obtain steady state solutions by solving the time evolution of these equations.
Without high-energy particles, we obtain advection dominated solutions that have features consistent with those obtained by previous studies.
Including high-energy particles, we also obtains advection dominated flows, 
and effects of high-energy particles on the flow structure turn out to be small 
even if the pressure of high-energy particles dominates over the thermal pressure. 
For a model in which escape of high-energy protons takes away almost all energy, 
the accretion flow has the Keplerian angular momentum, slow infall velocity, and large surface density. 
However, this solution is inconsistent in point of ignoring the electron component. 
Thus, if HEPs affect the dynamical structure, electrons are expected to be important. 
%%If some mechanisms that weaken compressional heating, such as outflows, 
%%high-energy particles are expected to affect the flow structure more strongly. 

We calculate luminosities of escaping particles for these steady solutions. 
For small mass accretion rates and large injection Lorentz factors of high energy particles with large diffusion coefficients, 
the luminosity of diffusively escaping protons amounts to $\Lp\sim 10^{-2} \dot M c^2$. 
In contrast, for large mass accretion rates, 
the luminosity of escaping neutrons, $\Ln$, is larger than $\Lp$, 
and its maximum value is nearly the same as that of the protons $\Ln\sim 10^{-2} \dot M c^2$. 
The luminosities of gamma-rays and neutrinos are a few times larger than $\Ln$. 
We note that radiative processes are expected to be important for large mass accretion rates. 
Though high-energy particles have little influence on dynamical structures, 
it is possible to extract some amount of energy through high-energy particles. 
They are considered to play some roles for production of relativistic jets in terms of the mass and energy injections.

%%If the internal energy or kinetic energy of the thermal component is available for acceleration of non-thermal particle, 
%%$$$\Ln$ and $\Lp$ could be as large as $\Ljet$. 
%%Thus, in this context, the neutron escape cannot become the main source of the production of the relativistic jets. 

%%%%%%%
%%Acknowledgments
%%%%%%%%%%%%%

We thank the referee for useful comments. 
S.S.K. thanks T. Tsuribe for useful discussion about calculation methods. 
S.S.K. thanks K. Nagamine, Y. Fujita, and H. Tagoshi for continuous encouragement. 
This work is partly supported by Grant-in-Aid for JSPS Fellows No. 231446 (K.T.) and No. 251784 (S.S.K.).



\appendix
\section{Derivation of the compressional heating term} \label{app:comp}
Equation (\ref{eq:eoc}) can be expressed as  
\begin{equation}
  \frac{d\ln \Sigma}{dt} = -\frac{1}{r}\delr (rv_r),
  \label{eq:dsigma}
\end{equation}
where $d/dt$ is the Lagrangian derivative. 
The equation of continuity is written as 
\begin{equation}
  \frac{d\ln \rho}{dt} = -\mbox{\boldmath $\nabla\cdot v$}
\end{equation}
Using $\rho = \Sigma/(2H)$, we obtain
\begin{equation}
   \mbox{\boldmath $\nabla\cdot v$} = -\frac{d \ln \rho}{dt}
  = -\frac{d \ln \Sigma}{dt} + \frac{d \ln H}{dt}\\
 = \frac{1}{r}\delr (rv_r) + \frac{d \ln H}{dt}
  \label{eq:divv2}
\end{equation}
Since $\Sigma$ and $H$ are independent of $z$, we find that $\mbox{\boldmath $\nabla\cdot v$}$ is independent of $z$. 
Thus, we can integrate the compressional heating term as 
\begin{eqnarray}
 \Qcompth = - \int \pth  (\mbox{\boldmath $\nabla\cdot v$}) dz
 = -\frac{\Pth}{r}\delr (rv_r) - \frac{\Pth v_r}{H} \delr H \\
  =  -\frac{\Pth}{rH}\delr (rHv_r). 
\end{eqnarray}
Here, we assume that $\partial/\partial t = 0 $ since we are interested in the steady solutions. 
This expression of $\mbox{\boldmath $\nabla\cdot v$}$ is the same as that of \citet{le05} though its derivation is a little different. 
We can derive the compressional heating term for HEPs in the same way by replacing $\pth$ with $\phe$.

\clearpage

\begin{table}
\begin{center}
\caption{Fixed parameters \label{tab:fix}}
\begin{tabular}{cccc}
\tableline\tableline
$\mbh/M_{\odot}$ &$\rout/\rs$ & $\rin/\rs$ & $\rcrit/\rs$ \\
\tableline
$10^8$  &150 & 1.5  & 100  \\
\tableline
\end{tabular}
%% Any table notes must follow the \end{tabular} command.
\end{center}
\end{table}

\begin{table}
\begin{center}
\caption{models and their parameters \label{tab:models}}
\begin{tabular}{cccccccc}
\tableline\tableline
 models & $\alpha$ & $\beta$ & $\dot M/\MdEdd$ & $\fv$ & $\fc$ & $\gammainj$ & $\Cdiff$\\
\tableline
 A1 & 0.1& 10 & 0.01 & 0.0 & 0.0 & -- & --  \\
 A2 & 0.1& 3 & 0.01 & 0.0 & 0.0 & -- & --  \\
 A3 & 0.1 & 10 & 0.001 & 0.0 & 0.0 & -- & --  \\
 A4 & 0.003 & 10 & 0.01 & 0.0 & 0.0 & -- & --  \\
\tableline
 B1 & 0.1 & 10 & 0.01 & 0.3 & 0.0 & $10^3$ & $10^4$  \\
 B2 & 0.1 & 10 & 0.01 & 0.9 & 0.0 & $10^3$ & $10^4$  \\
\tableline
 C1 & 0.1 & 10 & 0.01 & 0.0 & 0.3 & $10^3$ & $10^4$  \\
 C2 & 0.1 & 10 & 0.01 & 0.0 & 0.9 & $10^3$ & $10^4$  \\
\tableline
 D1 & 0.1 & 10 & 0.01 & 0.3 & 0.3 & $10^3$ & $10^4$  \\
 D2 & 0.1 & 10 & 0.01 & 0.6 & 0.6 & $10^3$ & $10^4$  \\
 D3 & 0.1 & 10 & 0.01 & 0.9 & 0.9 & $10^3$ & $10^4$  \\
\tableline
 E1 & 0.1 & 10 & 0.01 & 0.9 & 0.9 & $10^3$ & $10^6$  \\
\tableline
\end{tabular}
%% Any table notes must follow the \end{tabular} command.
\end{center}
\end{table}

\begin{table}
\begin{center}
\caption{models and their parameters \label{tab:dependence}}
\begin{tabular}{cccccccc}
\tableline\tableline
 models & $\alpha$ & $\beta$ & $\dot M/\MdEdd$ & $\fv$ & $\fc$ & $\gammainj$ & $\Cdiff$\\
\tableline
 F1 & 0.1 & 3 & 0.01 & 0.3 & 0.3 & $10^3$ & $10^4$  \\
 F2 & 0.1 & 3 & 0.01 & 0.6 & 0.6 & $10^3$ & $10^4$  \\
 F3 & 0.1 & 3 & 0.01 & 0.9 & 0.9 & $10^3$ & $10^4$  \\
\tableline
 G1 & 0.1 & 10 & 0.0001 & 0.3 & 0.3 & $10^3$ & $10^4$  \\
 G2 & 0.1 & 10 & 1.0 & 0.3 & 0.3 & $10^3$ & $10^4$  \\
\tableline
 H1 & 0.1 & 10 & 0.01 & 0.3 & 0.3 & $10^3$ & $10^2$  \\
 H2 & 0.1 & 10 & 0.01 & 0.3 & 0.3 & $10^3$ & $10^6$  \\
\tableline
 I1 & 0.1 & 3 & 0.01 & 0.3 & 0.3 & $10^3$ & $10^2$  \\
 I2 & 0.1 & 3 & 0.01 & 0.3 & 0.3 & $10^3$ & $10^4$  \\
 I3 & 0.1 & 3 & 0.01 & 0.3 & 0.3 & $10^3$ & $10^6$  \\
\tableline
 J1 & 0.1 & 10 & 0.01 & 0.3 & 0.3 & $10^1$ & $10^4$  \\
 J2 & 0.1 & 10 & 0.01 & 0.3 & 0.3 & $10^5$ & $10^4$  \\
\tableline
 K1 & 0.1 & 3 & 0.01 & 0.3 & 0.3 & $10^1$ & $10^4$  \\
 K2 & 0.1 & 3 & 0.01 & 0.3 & 0.3 & $10^3$ & $10^4$  \\
 K3 & 0.1 & 3 & 0.01 & 0.3 & 0.3 & $10^5$ & $10^4$  \\
\tableline

\end{tabular}
%% Any table notes must follow the \end{tabular} command.
\end{center}
\end{table}

% %% Use the figure environment and \plotone or \plottwo to include
% %% figures and captions in your electronic submission.
% %% To embed the sample graphics in
% %% the file, uncomment the \plotone, \plottwo, and
% %% \includegraphics commands
% %%
% %% If you need a layout that cannot be achieved with \plotone or
% %% \plottwo, you can invoke the graphicx package directly with the
% %% \includegraphics command or use \plotfiddle. For more information,
% %% please see the tutorial on "Using Electronic Art with AASTeX" in the
% %% documentation section at the AASTeX Web site, http://aastex.aas.org/
% %%
% %% The examples below also include sample markup for submission of
% %% supplemental electronic materials. As always, be sure to check
% %% the instructions to authors for the journal you are submitting to
% %% for specific submissions guidelines as they vary from
% %% journal to journal.

% %% This example uses \plotone to include an EPS file scaled to
% %% 80% of its natural size with \epsscale. Its caption
% %% has been written to indicate that additional figure parts will be
% %% available in the electronic journal.

% \begin{figure}
% \epsscale{.80}
% \plotone{f1.eps}
% \caption{Derived spectra for 3C138 \citep[see][]{heiles03}. Plots for all sources are available
% in the electronic edition of {\it The Astrophysical Journal}.\label{fig1}}
% \end{figure}

 \clearpage

% %% Here we use \plottwo to present two versions of the same figure,
% %% one in black and white for print the other in RGB color
% %% for online presentation. Note that the caption indicates
% %% that a color version of the figure will be available online.
% %%

% \begin{figure}
% \plottwo{f2.eps}{f2_color.eps}
% \caption{A panel taken from Figure 2 of \citet{rudnick03}. 
% See the electronic edition of the Journal for a color version 
% of this figure.\label{fig2}}
% \end{figure}

% %% This figure uses \includegraphics to scale and rotate the still frame
% %% for an mpeg animation.

\begin{thebibliography}{}
\bibitem[Abramowicz et al.(1988)]{abr88} 
	Abramowicz, M. A., Czerny, B., Lasota, J. P., \& Szuszkiewicz, E. 1988, \apj, 332, 646
\bibitem[Aharonian \& Atoyan(2000)]{aha00} 
	Aharonian, F. A., \& Atoyan, A. M. 2000, \aap, 362, 937
\bibitem[Asano \& Takahara(2007)]{asa07} 
	Asano, K., \& Takahara, F. 2007, ApJ, 655, 762
\bibitem[Balbus \& Hawley(1991)]{bal91} 
	Balbus, S. A., \& Hawley, J. F. 1991, \apj, 376, 214
\bibitem[Becker et al.(2008)]{bec08} 
	Becker, P. A., Das, S., \& Le, T. 2008, \apj, 677, L93
\bibitem[Becker et al.(2011)]{bec11} 
	Becker, P. A., Das, S., \& Le, T. 2011, \apj, 743, 47
\bibitem[Begelman et al.(1990)]{beg90} 
	Begelman, M. C., Rudak, B., \& Sikora, M. 1990, \apj, 362, 38
\bibitem[Blandford \& Begelman(1999)]{bla99} 
	Blandford R. D. \& Begelman M. C., 1999, \mnras, 303, L1
\bibitem[Chen et al.(1997)]{che97} 
	Chen, X., Abramowicz, M., \& Lasota, J.-P. 1997, \apj, 476, 61
\bibitem[Dermer et al.(1996)]{der96} 
	Dermer, C. D., Miller, J. A., \& Li, H. 1996, ApJ 456, 106
\bibitem[Drury(1983)]{dru83} 
	Drury, L. O'C., 1983, Rep. Progress Phys., 46, 973
\bibitem[Esin et al.(1997)]{esi97} 
	Esin, A. A., McClintock, J. E., \& Narayan, R. 1997, ApJ, 489, 865
\bibitem[Fernandes et al.(2011)]{fer11} 
	Fernandes, C. A. C., Jarvis, M. J., Rawlings, S., et al. 2011, MNRAS, 411, 1909 
\bibitem[Hoshino(2013)]{hos13} 
	Hoshino, M. 2013, \apj, 773, 118 
\bibitem[Inoue \& Takahara(1996)]{ino96} 
	Inoue, S. \& Takahara, F. 1996, \apj, 463, 555
\bibitem[Jones(1990)]{jon90} 
	Jones, F. 1990, \apj, 361, 162
\bibitem[Katz(1991)]{kat91} 
	Katz, J. I. 1991, ApJ, 367, 407
\bibitem[Kato et al.(2008)]{kat08} 
	Kato, S., Fukue, J., \& Mineshige, S. 2008, Black-Hole Accretion Disks-Towards a New Paradigm (Kyoto: Kyoto Univ. Press)
\bibitem[Komissarov et al.(2007)]{kom07} 
	Komissarov, S. S., Barkov, M. V., Vlahakis, N., \& Ko ̈nigl, A. 2007, MNRAS, 380, 51
\bibitem[Le \& Becker(2004)]{le04} 
	Le, T., \& Becker, P. A. 2004, \apj, 617, L25
\bibitem[Le \& Becker(2005)]{le05} 
	Le, T., \& Becker, P. A. 2005, \apj, 632, 476
\bibitem[Mahadevan et al.(1997)]{mah97a} 
	Mahadevan R., Narayan R., \& Krolik J., 1997, ApJ, 486, 268 
\bibitem[Mahadevan \& Quataert(1997)]{mah97b} 
	Mahadevan, R., \& Quataert, E. 1997, ApJ, 490, 605
\bibitem[Manmoto et al.(1997)]{man97} 
	Manmoto, T., Mineshige, S., \& Kusunose M. 1997, ApJ, 489, 791
\bibitem[Matsumoto et al.(1984)]{mat84} 
	Matsumoto, R., Fukue, J., Kato, S., \& Okazaki, A. S. 1984, PASJ, 36, 71
\bibitem[McKinney(2006)]{mck06} 
	McKinney, J. C. 2006, \mnras, 368, 1561
\bibitem[Nakamura et al.(1997)]{nak97} 
	Nakamura, E. K., Kusunose, M., Matsumoto, R., \& Kato, S. 1997, \pasj, 49, 503
\bibitem[Narayan \& Yi(1994)]{nar94} 
	Narayan, R., \& Yi, I. 1994, \apj, 428, L13
\bibitem[Narayan \& Yi(1995)]{nar95} 
	Narayan, R., \& Yi, I. 1995, \apj, 452, 710
\bibitem[Narayan et al.(1997)]{nar97} 
	Narayan R., Kato S., Honma F., 1997 ApJ, 476, 49
\bibitem[Niedzwiecki et al.(2013)]{nie13} 
	Nied\'{z}wiecki, A., Xie, F.-G., \& Stepnik, A. 2013, MNRAS, 432, 1576
\bibitem[Oda et al.(2007)]{oda07} 
	Oda, H., Machida, M., Nakamura, K., \& Matsumoto, R. 2007, \pasj, 59, 457
\bibitem[Ohsuga \& Mineshige.(2011)]{ohs11} 
	Ohsuga, K. \& Mineshige, S. 2011, \apj, 736, 2
\bibitem[Paczy\'{n}sky \& Wiita (1980)]{pac80} 
	Paczy\'{n}ski B. \& Wiita P. J., 1980, \aap, 88, 23
\bibitem[Press et al.(1992)]{pre92} 
	Press, W. H., Teukolsky, S. A., William, T. V., \& Brian, P. F. 1992, NUMERICAL RECIPES in FORTRAN (2nd ed.; Cambridge University press)
\bibitem[Pringle (1981)]{pri81} 
	Pringle J.E., 1981, ARA\&A, 19, 137
\bibitem[Punsly \& Zhang (2011)]{pun11} %
	Punsly, B., \& Zhang, S. 2011, ApJ, 735, L3
\bibitem[Riquelme et al.(2012)]{riq12} 
	Riquelme, M. A., Quataert, E., Sharma, P., \& Spitkovsky, A. 2012, \apj, 755, 50 
\bibitem[Sano et al.(2004)]{san04} 
	Sano, T., Inutsuka, S., Turner, N. J., \& Stone, J. M. 2004, \apj, 605, 321
\bibitem[Shakura \& Sunyaev(1973)]{sha73} 
	Shakura, N. I., \& Sunyaev, R. A. 1973, \aap, 24, 337
\bibitem[Spitzer(1962)]{spi62} 
	Spitzer, L. 1962. Physics of Fully Ionized Gases, pp. 120–154. Interscience, New York.
\bibitem[Stone \& Norman(1992)]{sto92} 
	Stone, J. M., \& Norman, M. L. 1992, ApJS, 80, 753
\bibitem[Subramanian et al.(1999)]{sub99} 
	Subramanian, P., Becker, P. A., \& Kazanas, D. 1999, ApJ, 523, 203 
\bibitem[Takahara \& Kusunose(1985)]{tak85} 
	Takahara, F., \& Kusunose, M. 1985, Prog. Theor. Phys., 73, 1390
\bibitem[Toma \& Takahara(2012)]{tom12} 
	Toma, K. \& Takahara, T. 2012, \apj, 754, 148
\bibitem[von Neumann \& Richtmyer(1950)]{neu50} 
	von Neumann, J. \& Richtmyer, R. D. 1950, Appl. Phys., 21, 232
\bibitem[Yuan(2001)]{yua01} 
	Yuan, F., 2001, \mnras, 327, 119


 \end{thebibliography}
\end{document}